\documentclass[12pt,a4paper]{article}
\usepackage{amssymb,mathrsfs}
\usepackage{amsmath,amssymb,latexsym,amsthm,color,times,fullpage}
\usepackage{booktabs,graphicx}
\usepackage{tikz,pifont,bm}
\usepackage[pdftex]{hyperref}
\hypersetup{plainpages=True, pdfstartview=FitV,
colorlinks=true,linkcolor=blue,citecolor=blue}

\def\v#1{{\mathbf{#1}}}
\def\mb#1{{\mathbb{#1}}}
\def\tr#1{\left\lfloor#1\right\rfloor}
\newtheorem{thm}{Theorem}
\newtheorem{lmm}{Lemma}
\newtheorem{cor}{Corollary}
\newtheorem{prop}{Proposition}
\def\pf{\noindent \emph{Proof.}\ }
\def\qed{{\quad\rule{3mm}{3mm}\,}}

\title{Threshold phenomena in $k$-dominant skylines of
random samples}
\author{ {\sc Hsien-Kuei Hwang}\\
    Institute of Statistical Science\\
    Academia Sinica\\
    Taipei 115\\
    Taiwan
\and {\sc Tsung-Hsi Tsai}\\
    Institute of Statistical Science\\
    Academia Sinica\\
    Taipei 115\\
    Taiwan
\and {\sc Wei-Mei Chen}\\
    Department of Electronic Engineering\\
    National Taiwan University of Science and Technology\\
    Taipei 106\\
    Taiwan
}
\date{\today}
\begin{document}
\maketitle

\begin{abstract}

Skylines emerged as a useful notion in database queries for
selecting representative groups in multivariate data samples for
further decision making, multi-objective optimization or data
processing, and the $k$-dominant skylines were naturally introduced
to resolve the abundance of skylines when the dimensionality grows
or when the coordinates are negatively correlated. We prove in this
paper that the expected number of $k$-dominant skylines is
asymptotically zero for large samples when $1\le k\le d-1$ under two
reasonable (continuous) probability assumptions of the input points,
$d$ being the (finite) dimensionality, in contrast to the asymptotic
unboundedness when $k=d$. In addition to such an asymptotic
zero-infinity property, we also establish a sharp threshold
phenomenon for the expected ($d-1$)-dominant skylines when the
dimensionality is allowed to grow with $n$. Several related issues
such as the dominant cycle structures and numerical aspects, are
also briefly studied.

\end{abstract}

\noindent \emph{Key words.} Skyline, dominance, maxima, random
samples, Pareto optimality, threshold phenomena, multi-objective
optimization, computational geometry, asymptotic approximations,
average-case analysis of algorithms.

\section{Introduction}

The last decade has undergone a drastic change of information
dissemination from Web 1.0 to Web 2.0, the most notable
representative products being YouTube and Facebook. Data have been
generated in an unprecedented pace and range, powerful search
engines are indispensable, and screening useful or usable
information (via ``sort engines'') from the vast is generally
becoming more important than searching and gathering. Skylines of
multivariate data sample were introduced for selecting
representative groups in the database query literature by
B\"{o}rzs\"{o}nyi et al.\ (see \cite{BKS01}) and had appeared in
diverse areas under several different guises and names: \emph{Pareto
optimality}, \emph{efficiency}, \emph{maxima}, \emph{admissibility},
\emph{elite}, \emph{sink}, etc.; see \cite{CHT03,CHT11} and the
references therein for more information. These diverse terms reveal
the importance of the use of skyline as an effective means of data
summarization in theory and in practice. Many different notions and
variants of skylines have been proposed in the literature, following
the original paper \cite{BKS01}. In particular, the $k$\!-dominant
skylines were introduced by Chan et al.\ (see \cite{CJTTZ06}) in
situations when the skylines are abundant and have received much
attention since, although they had already been studied in the
Russian literature (see for example \cite{BO96,Orlova91}). We focus
in this paper on the asymptotic estimates of such skylines and prove
several types of threshold phenomena under different probability
assumptions of the input samples, which, in addition to their
theoretical interests, are believed to be useful for practitioners.

\paragraph{Skylines and $k$-dominant skylines} The definitions of
skyline and many of its variants are based on the notion of
dominance. Given a $d$-dimensional dataset $\mathscr{D}$, a point
$\v{p}\in\mathscr{D}$ is said to \emph{dominate} another point
$\v{q}\in\mathscr{D}$ if $p_j\le q_j$ for $1\le j\le d$, where
$\v{p}=(p_1,\dots,p_n)$ and $\v{q}=(q_1,\dots,q_n)$, and is less
than in at least one dimension. The non-dominated points in
$\mathscr{D}$ are called the \emph{skyline} (or \emph{skyline
points}) of $\mathscr{D}$. By relaxing the full dominance definition
to partial dominance, we say that a point $\v{p}\in\mathscr{D}$
\emph{$k$-dominates} another point $\v{q} \in\mathscr{D}$ if there
are $k$ dimensions in which $p_j$ is not greater than $q_j$ and is
less than in at least one of these $k$ dimensions\footnote{If we
change the definition of the $k$-dominant skyline to be ``exactly
$k$'' (instead of $\ge k$) coordinates smaller than or equal to and
at least $1$ smaller than, then the same types of results in this
paper also hold.}. The points in $\mathscr{D}$ that are not
$k$-dominated by any other points are defined to be the
\emph{$k$-dominant skyline} of $\mathscr{D}$; see \cite{CJTTZ06}.
See also \cite{BO96} for a different formulation.

The definition of $k$-dominant skyline implies that for a fixed
dataset the number of $k$-dominant skylines decreases as $k$ becomes
smaller. Such a monotonicity property will be used later. To see
this, consider any point $\v{p}$ in the unit square. It is a skyline
(or $2$-dominant skyline) point if no other points have
simultaneously smaller $x$- and smaller $y$-values; namely, no other
points can lie in the shaded region
\begin{tikzpicture}[scale=0.3]
\begin{scope}
\clip (-0.5,-0.5) rectangle (0,0);%
\foreach \k in {0.8,0.7,...,-0.8} {%
\draw [red!40,rotate=-40](\k,-2) -- (\k,2);%
}%
\end{scope}
\draw[-latex,line width=0.3pt](-0.5,-0.5) rectangle (0.5,0.5);
\draw[line width=0.3pt] (0,-0.5) -- (0,0.5);%
\draw[line width=0.3pt] (-0.5,0) -- (0.5,0);%
\filldraw[] circle(0.07);%
\end{tikzpicture} (where $\v{p}$ is the dotted point in the middle
of this figure). However, to be a $1$-dominant skyline point
requires that all other points must have simultaneously larger $x$-
and larger $y$-values, or, equivalently, they cannot lie in the
shaded region
\begin{tikzpicture}[scale=0.3]
\begin{scope}
\clip (-0.5,0) rectangle (0,0.5);%
\foreach \k in {0.8,0.7,...,-0.8} {%
\draw [red!40,rotate=-40](\k,-2) -- (\k,2);%
}%
\end{scope}
\begin{scope}
\clip (-0.5,-0.5) rectangle (0.5,0);%
\foreach \k in {0.8,0.7,...,-0.8} {%
\draw [red!40,rotate=-40](\k,-2) -- (\k,2);%
}%
\end{scope}
\draw[-latex,line width=0.3pt](-0.5,-0.5) rectangle (0.5,0.5);
\draw[line width=0.3pt] (0,-0.5) -- (0,0.5);%
\draw[line width=0.3pt] (-0.5,0) -- (0.5,0);%
\filldraw[] circle(0.07);%
\end{tikzpicture}.

On the other hand, the transitivity property of skylines fails for
$k$-dominant skylines when $1\le k\le d-1$, meaning that their
cardinality may be zero and there may be cycles.

\paragraph{The number of skyline points} The number of skyline
points is a key issue in their use and usefulness. This quantity
under suitable random assumptions of the input is also important for
practical modeling or reference purposes, as well as for the
analysis of skyline-finding algorithms. The two major, simple,
representative random models are \emph{hypercubes} and
\emph{simplices}. Assuming that the input dataset $\mathscr{D}
=\{\v{p}_1,\dots, \v{p}_n\}$ is taken uniformly and independently
from the hypercube $[0,1]^d$, then it has been known since the
1960's (see \cite{BNS66}) that the expected number of skyline points
of $\mathscr{D}$ is asymptotic to $\frac{(\log n)^{d-1}}{(d-1)!}$
for large $n$ and finite $d$, exhibiting the independence of the
coordinates. (Intuitively, if one sorts according to one dimension,
then each other dimension roughly contributes $\log n$ skyline
points.) On the other hand, if we assume that the input points are
uniformly sampled from the $d$-dimensional simplex $\{|x_1|+\cdots
+|x_d|\le 1, x_j\in(-1,0]\}$, then the expected number of skyline
points is asymptotic to $\Gamma \left(\frac1d\right) n^{1-\frac1d}$,
reflecting obviously a stronger negative correlation of the
coordinates; see \cite{BDHT05} and the references cited there. Here
$\Gamma$ denotes Euler's Gamma function. For the number of skyline
points under other models, see \cite{Baryshnikov07, Devroye86,
Devroye93,SY08} and the references therein.

On the other hand, in contrast to the recent growing trend of
studying high dimensional datasets, not much is known for the
expected number of skyline points when $d$ is allowed to grow with
$n$. Such a direction is especially useful as practical situations
always deal with finite $n$ and finite $d$ (whose dependence on $n$
is often not clear). The only exception along this direction is the
uniform estimates given in \cite{Hwang04} (see also \cite{BDHT05})
for the expected number of skyline points in a random uniform
samples of $n$ points from the hypercube $[0,1]^d$. While the order
$\frac{(\log n)^{d-1}} {(d-1)!}$ may seem slowly growing as $d$
increases, it soon reaches the order $n$ when $d$ is around $\log
n$, which is relatively small for moderate values of $n$.
Consequently, the skyline points become too numerous to be of direct
use. The growth of skyline points in the random $d$-dimensional
simplex model is even faster and we can show that almost all points
are skylines when $d$ roughly exceeds $\frac{\log n} {\log\log n}$,
again small for $n$ not too large.

\paragraph{The cardinality of $k$-dominant skyline} Since
$k$-dominant skyline were proposed (see \cite{CJTTZ06}) to resolve
the skyline-abundance problem, it is of interest to know their
quantity under suitable random models. A critical step in applying
$k$-dominant skyline is to identify an appropriate $k$ such that the
size of the $k$-dominant skyline is within the acceptable ranges.
But this may not be always feasible. Consider the $5$-dimensional
dataset $\mathscr{D}$ given in Table~\ref{tab:ex}. The six points
are all skyline points, one ($\v{p}_6$) is the $4$-dominant skyline
point and no point is in the $3$-dominant skyline. Clearly,
$\v{p}_6$ is to some extent better than the other points since it
contains two components with the lowest value $1$. However, it was
already mentioned in \cite{CJTTZ06} that some $k$-dominant skylines
may be empty. For example, if we drop $\v{p}_6$ from $\mathscr{D}$,
then the five points are all skyline points but all $k$-dominant
skylines are empty for $1\le k\le 4$. In this example, other
alternatives to $k$-dominant skylines have to be used.
Unfortunately, such a property of \emph{excessive skylines but few
$k$-dominant skylines} is not uncommon, and we show in this paper
that, under the hypercube and the simplex random models, the
expected number of $k$-dominant skylines both tends to zero for
large $n$ and $1\le k\le d-1$.

\medskip
\begin{center}
\label{tab:ex}
\begin{tabular}{c|cccc}\toprule
point  & skyline& $4$-dominant skyline & $3$-dominant skyline\\
\midrule
$\v{p}_1{\ }(1,2,2,3,3)$&\ding{52}&-&-\\
$\v{p}_2{\ } (3,1,2,2,3)$&\ding{52}&-&-\\
$\v{p}_3{\ } (3,3,1,2,2)$&\ding{52}&-&-\\
$\v{p}_4{\ } (2,3,3,1,2)$&\ding{52}&-&-\\
$\v{p}_5{\ } (2,2,3,3,1)$&\ding{52}&-&-\\
$\v{p}_6{\ } (2,3,1,1,3)$&\ding{52}&\ding{52}&-\\
\bottomrule
\end{tabular}\vspace*{.5cm}
\centerline{\emph{Table 1: An example showing the property of
many skylines but few $k$-dominant skylines.}}
\end{center}
\medskip

\paragraph{Threshold phenomena} We clarify two types of threshold
phenomena for the expected number of $k$-dominant skylines in random
samples.
\begin{enumerate}

\item \emph{Large sample, bounded dimension}:
\[
    \text{Expected number of $k$-dominant skylines}
    \to \left\{\begin{array}{ll}
        0, & \text{if }1\le k\le d-1;\\
        \infty,& \text{if }k=d,
    \end{array}
    \right.
\]
as the sample size $n\to\infty$. While such a result is not new and
contained as a special case of the general theory developed in
\cite{BO96} for finite dimensional skylines, we will give an
independent, transparent, self-contained proof, which, in addition
to being more precise, can be extended to the case when the
dimensionality goes unbounded with the sample size.

\item \emph{Large sample, moderate dimension}: There exists
an integer $d_0=d_0(n)\approx \sqrt{\frac{2\log n}{\log\frac{\log n}
{\log\log n}}} +1$ such that (see \eqref{EMkn-large})
\begin{align*}
    \text{Expected number of $(d-1)$-dominant skylines}
    \to \left\{\begin{array}{ll}
        0, &\text{if } d\le d_0-1;\\
        \infty,& \text{if } d\ge d_0+2,
    \end{array} \right.
\end{align*}
as $n\to\infty$, and the two cases $d=d_0$ and $d=d_0+1$ lead to
two different oscillating functions, the first ($d=d_0$)
fluctuating between $0$ and $\frac{e^{-\gamma}}{2-e^{-e^{-1}}}$
and the second between $\frac{e^{-\gamma}}{2-e^{-e^{-1}}}$ and
$O\left(\frac{\log n}{\log\log n}\right)$, 
where $\gamma$ is Euler's constant; see \eqref{n-ii2}
and \eqref{n-ii3}. We consider only random samples from
hypercubes. Other regions and other values of $k$, $k<d-1$ are
expected to exhibit similar threshold phenomena with different
$d_0$, but the analysis becomes excessively long and involved.
More details will be discussed elsewhere.
\end{enumerate}
We see from these phenomena that the usual ``curse of high
dimensionality'' has thus another form here which one may term
``curse of constant dimensionality,'' which refers to the situation
when no $k$-dominant skyline point at all exists. Also the model
where dimensionality can vary with the sample size is, at least from
a practical point of view, more reasonable; see
Sections~\ref{sec:d-large} and \ref{sec:threshold} for more
discussions and details.

\paragraph{Related works} In addition to the partial dominance
used in defining $k$-dominant skylines (see \cite{CJTTZ06}), there
are also several other skyline variants for retrieving more
representative points; these include skybands \cite{PTFS05}, top-$k$
dominating queries \cite{IBS08,PTFS05,YM09}, strong skylines
\cite{ZGLTW05}, skyline frequency \cite{CJTTZ062}, approximately
dominating representatives \cite{KP07}, $\varepsilon$-skylines
\cite{XZT08}, and top-$k$ skylines \cite{BGG07, LYH09}. See also the
survey paper \cite{IBS08} for more information.

\paragraph{Organization of the paper} This paper presents a
systematic study on the asymptotic estimates of the number of
$k$-dominant skyline points under random models. It is organized as
follows. We derive in the next section (\S~\ref{sec:1st}) an
asymptotic vanishing property for the number of $k$-dominant skyline
points under a common hypercube model when the dimensionality is
bounded. The extension to include more points in the partial
dominant skyline is showed to suffer from a similar drawback in
Section~\ref{sec:layers}. We then prove in Section~\ref{sec:simplex}
that changing the underlying model from hypercube to simplex does
not improve either the asymptotic vanishing property.
Section~\ref{sec:cm} deals with a categorical model for which the
results have a very different nature. Roughly, as the total number
of sample points are finite in this model, the expected number of
$k$-dominant skylines will be asymptotically linear, meaning too
many choices for ranking or selection purposes. All these results
point to the negative side for the use of $k$-dominant skylines
under similar data situations. We then address the positive side in
the last few sections by considering again the hypercubes but with
growing dimensionality. A sharp threshold phenomenon is discovered
in Section~\ref{sec:threshold} when $d\to\infty$ with $n$, the
asymptotic approximations needed being derived in
Section~\ref{sec:d-large}. Another new threshold result is given in
Section~\ref{sec:nc} of the expected number of dominant cycles.
Section~\ref{sec:ub} provides a uniform lower-bound estimate for the
expected number of skyline points for $1\le k\le d-1$. We conclude
in Section~\ref{sec:fin} with some numerical aspects of the
estimates we derived.

\section{Random samples from hypercubes}
\label{sec:1st}

The simplest random model is the hypercube $[0,1]^d$, which is also
the most natural and most studied one. They can also be used when
data are discrete in nature but span uniformly over a sufficiently
large interval.

In this section, we derive asymptotic estimates for the expected
number of $k$-dominant skyline points in a random sample of $n$
points $\mathscr{D} := \{\v{p} _{1},\ldots ,\v{p}_n\}$ uniformly and
independently drawn from $[0,1]^d$, $d\ge2$. Let $M_{d,k}(n)$ denote
the number of $k$-dominant skyline points of $\mathscr{D}$. We first
derive a crude upper bound for the expected number
$\mathbb{E}[M_{d,k}(n)]$, which implies that
$\mathbb{E}[M_{d,k}(n)]$ is asymptotically zero as $n$ grows
unbounded and $1\le k\le d-1$. More precise estimates are possible
and will be derived in Section~\ref{sec:d-large}. For a point
$\v{p}\in [0,1]^d$, denoted by $B_{k}(\v{p})$ the region of the
points in $[0,1]^d$ that $k$-dominates $\v{p}$. Also, $|A|$ denotes
the volume of the region $A$.

\begin{thm}[Asymptotic zero-infinity property for large $n$ and
bounded $d$] For fixed $d\ge2$ \label{thm-q1}
\begin{equation}\label{a1}
    \mathbb{E}[M_{d,k}(n)]
    \to \left\{\begin{array}{ll}
        0, & \text{if }1\le k\le d-1;\\
        \infty,& \text{if }k=d,
    \end{array}
    \right.
\end{equation}
as $n\to\infty$.
\end{thm}
\pf  The case $k=d$ has been known since the 1960's (see
\cite{BNS66}) and were re-derived several times in the literature. 
We assume $1\le k \le d-1$. Since $M_{d,k}(n)\le
M_{d,d-1}(n)$ for fixed $d$ and for $1\le k\le d-1$, we only prove
that $\mathbb{E}[M_{d,d-1}(n)]\to 0$.

We start from the integral representation
\begin{align}
    \mathbb{E}[M_{d,d-1}(n)] &= n\mb{P}\left(\v{p}_1 \text{ is a
    ($d-1$)-dominant skyline point}\right) \nonumber \\
    &= n\int_{[0,1]^d} \left(1-|B_{d-1}(\v{x})|\right)^{n-1}
    \text{d}\v{x}\label{EM} ,
\end{align}
because if $\v{x}$ is not $k$-dominated by any of the other $n-1$
points, they all have to lie in the region $[0,1]^d \setminus
B_k(\v{x})$. Here and throughout this paper, $\text{d}\v{x}$ is the
abbreviation of $\text{d}x_1\cdots \text{d}x_d$.

To estimate the integral in \eqref{EM}, we split it into two parts,
one part having sufficiently small volume (corresponding roughly to
small $x_1\cdots x_d$) and the other with $|B_{d-1}(\v{x})|$ bounded
away from zero, rendering the term $(1-|B_{d-1}(\v{x})|)^{n-1}$ also
small.

For a fixed number $t$ satisfying $1<t<\frac{d}{d-1}$, define the
region
\begin{align} \label{Qn}
    Q_{n}
    :=\bigcup_{1\le \ell \le d}\left\{\v{x}\in [0,1]^{d}:
    x_{\ell}\le n^{-\frac td}\text{ and }
    \prod_{j\not= \ell} x_j
    \le n^{-\frac{d-1}d\,t}\right\} .
\end{align}
Then
\[
    \mathbb{E}[M_{d,d-1}(n)]
    \le n\left| Q_{n}\right|+n\int_{[0,1]^{d}\setminus Q_{n}}
    \left( 1-\left| B_{d-1}(\v{x})\right|\right)^{n-1}
    \mathrm{d} \v{x}.
\]
The volume of $Q_n$ is bounded above by
\begin{align*}
    \left| Q_{n}\right|
    \le dn^{-\frac td} \int_{\substack{x_1\cdots x_{d-1}
    \le n^{-\frac{d-1}{d}\,t}
    \\ \v{x}\in[0,1]^d}}\text{d}\v{x}.
\end{align*}
To estimate the last integral, let
\[
    A_d(\delta) := \int_{\substack{x_1\cdots x_{d-1}\le \delta
    \\ \v{x}\in[0,1]^d}}\text{d}\v{x} \qquad(d\ge2),
\]
where $0<\delta<1$. Then $A_2(\delta) = \delta$, and
\[
    A_d(\delta) = \int_{\delta}^1 A_{d-1}
    \left(\frac{\delta}{t}\right)
    \text{d}t \qquad(d\ge3).
\]
A simple induction gives
\[
    A_d(\delta) = \delta\frac{|\log \delta|^{d-2}}{(d-2)!}
    \qquad(d\ge2),
\]
and we obtain, by taking $\delta = n^{-\frac{d-1}{d}\,t}$,
\[
    |Q_n|=O\left(n^{-t} (\log n)^{d-2}\right) ,\label{Qn-est}
\]

On the other hand, by an inclusion-exclusion argument, we have
\begin{align} \label{Bd}
    |B_{d-1}(\v{x})|
    =\sum_{1\le \ell \le d}\prod_{j\neq \ell}
    x_j-(d-1)\prod_{1\le j\le d}x_j.
\end{align}
Now if $\v{x}\in [0,1]^{d}\setminus Q_{n}$, then
\[
    \left| B_{d-1}(\v{x})\right|
    \ge \max_{1\le \ell \le d}\prod_{i\neq \ell }x_{i}
    \ge n^{-\frac{d-1}{d}\,t}.
\]
Thus, we have
\begin{align} \label{dm1-skl}
    \mathbb{E}[M_{d,d-1}(n)]
    = O\left( n^{1-t}(\log n)^{d-2}\right)
    +O\left( n\exp \left(-(n-1)n^{-\frac{d-1}{d}\,t}\right)\right),
\end{align}
and we see easily that the right-hand side tends to zero by our
choice of $t$. More precisely, if we take
\[
    t = \frac{d}{d-1}\left(1
    -\frac{\log\left(\frac d{d-1}\log n\right)}{\log n}\right),
\]
so as to balance the two $O$-terms in \eqref{dm1-skl}, then
\[
    \mathbb{E}[M_{d,d-1}(n)] = O\left(n^{-\frac1{d-1}}
    (\log n)^d\right).
\]
This and the monotonicity of $M_{d,k}(n)$ (in $k$) proves \eqref{a1}.
\qed

The fact that $\mathbb{E}[M_{d,k}(n)]\to0$ implies that there are
many cycles formed by the $k$-dominant relation, but the
corresponding cycle structures are very difficult to quantify; see
Section~\ref{sec:fin} for some preliminary results.

\section{``Clouds" of $k$-dominant skylines}
\label{sec:layers}

The asymptotic vanishing property (Theorem~\ref{thm-q1}) for the
expected number of $k$-dominant skylines limits their usefulness if
the input data are known to be in similar randomness conditions. In
particular, if one is interested in finding the top-$K$
representative points, then the probability of getting enough number
of candidates tends to zero. A simple remedy to this situation (and
still following the same notion of partial dominance between points)
is to consider the number of points that are $k$-dominated by a
specified number, say $j$ of other points, which we refer to as the
``cloud" of $k$-dominant skylines. But we show that this also
suffers from similar vanishing drawback under the random hypercube
model, unless $j$ is chosen to be large enough.

Let $L_{d,k}(n,j)$ denote the number of points in the random sample
$\{\v{p}_1,\ldots ,\v{p}_n\}$ that are $k$-dominated by exactly $j$
points, where the $n$ points are uniformly and independently
selected from $[0,1]^{d}$. Note that $L_{d,k}(n,0)$ is nothing but
$M_{d,k}(n)$.

\begin{thm}[Asymptotic zero-infinity property for clouds of
$k$-dominant skylines] For fixed $d\ge2$ and $1\le k\le d-1$,
\[
    \mathbb{E}[L_{d,k}(n,j)] \to \left\{\begin{array}{ll}
        0, & \text{if }1\le k\le d-1;\\
        \infty,& \text{if }k=d,
    \end{array}
    \right.
\]
uniformly for $0\le j=o(n^{(1-\varepsilon)/d})$, as $n\to\infty$,
where $\varepsilon>0$ is an arbitrarily small constant.
\end{thm}
The theorem roughly says that even allowing more flexible partial
dominance relation, the expected number of the skylines so
constructed still approaches zero as long as the dimensionality is
fixed.

\medskip

\pf  
The case when $k=d$ is also derived in \cite{BNS66} (under the
name of ``$(j+1)^{\text{st}}$ layer, 1-st quadrant-admissible
points"), where it is showed that
\[
    \mathbb{E}[L_{d,d}(n,j)] =
    \sum_{j<i_1\le\cdots\le i_{d-1}\le n}\frac1{i_1\cdots i_{d-1}},
\]
from which we obtain
\begin{align} \label{Lddnj}
    \mathbb{E}[L_{d,d}(n,j)]
    \sim \frac{\left(\log\frac{n}{j+1}\right)^{d-1}}{(d-1)!},
\end{align}
if $\log(n/(j+1))\to\infty$, where the symbol ``$\sim$" means that
the ratio of both sides tends to $1$ as $n$ goes unbounded.
Alternatively, we can use the integral representation (see
\cite{BCHL98})
\begin{align}
    \mathbb{E}[L_{d,d}(n,j)] &= n \binom{n-1}{j}\int_{[0,1]^d}
    \left(x_1\cdots x_d\right)^{j} \left(1-x_1\cdots x_d
    \right)^{n-1-j} \text{d} \v{x}\nonumber \\
    &= \frac{n}{(d-1)!}\binom{n-1}{j}\int_0^1
    t^j (1-t)^{n-1-j} \log\left(\tfrac1t\right)^{d-1}
    \text{d} t, \label{int-ELdd}
\end{align}
by the change of variables $t \mapsto x_1\cdots x_d$. A
straightforward evaluation then gives (\ref{Lddnj}).

Note that $\frac{\mathbb{E}[L_{d,d}(n,j)]}n$ equals the probability
that the first-quadrant subtree of the root has size $j$ in random
quadtrees; see \cite[Appendix]{FLLS95}. This connection also
provides several other expressions for $\mathbb{E}[L_{d,d}(n,j)]$.
For example,
\begin{align*}
    \mathbb{E}[L_{d,d}(n,j)]
    = \binom{n-1}{j}\sum_{0\le \ell\le n-1-j}\binom{n-1-j}{\ell}
    \frac{(-1)^\ell}{(j+1+\ell)^{d}};
\end{align*}
see also \cite{BDHT05}.

For the remaining cases, we consider only $k=d-1$ and prove that
$\mb{E}[L_{d,d-1}(n,j)]\to0$. The reason is that
\[
    \sum_{0\le \ell \le j} L_{d,k}(n,\ell)
    \le \sum_{0\le \ell \le j} L_{d,d-1}(n,\ell)
    \qquad(1\le k\le d-1).
\]
To see this, observe that if a point $\v{p}$ $(d-1)$-dominates
another point $\v{q}$, then $\v{p}$ also $k$-dominates $\v{q}$
for $1\le k\le d-2$. Thus, the sum on the left-hand side, which
stands for the set that is $k$-dominated by at most $j$ points,
is less than the sum on the right-hand side, the set that is
$(d-1)$-dominated by at most $j$ points.

To prove $\mb{E}[L_{d,d-1}(n,j)]\to0$, we apply the same argument
used in the proof of Theorem~\ref{thm-q1} starting from the integral
representation
\begin{align*}
    \mathbb{E}[L_{d,d-1}(n,j)]
    &= n\int_{[0,1]^d}\mb{P}(\text{exactly } j
    \text{ points in $\{\v{p}_2,\dots,\v{p}_n\}$
    that $k$-dominate } \v{p}_1)\\
    &= n\binom{n-1}{j}
    \int_{[0,1]^d} B_{d-1}(\v{x})^j
    \left(1-B_{d-1}(\v{x})\right)^{n-1-j}\text{d}\v{x}.
\end{align*}
Now we fix a constant $t$ satisfying $1<t<\frac d{d-1}$, and then
choose $Q_n$ as in \eqref{Qn}. Then we have
\[
    |Q_n| = O\left(n^{-t}(\log n)^{d-2}\right),
\]
and
\[
    n^{-\frac{d-1}{d}\,t} \le |B_{d-1}(\v{x})| \le 1
    \qquad(\v{x}\in[0,1]^d\setminus Q_n).
\]
It follows that
\begin{align*}
    \mathbb{E}[L_{d,d-1}(n,j)]
    &\le n|Q_n| + n\binom{n-1}{j}
    \int_{[0,1]\setminus Q_n} B_{d-1}(\v{x})^j
    \left(1-B_{d-1}(\v{x})\right)^{n-1-j}\text{d}\v{x}\\
    &= O\left(n^{1-t}(\log n)^{d-2}\right)
    + O\left(n\binom{n-1}{j} \exp\left(-(n-1-j)
    n^{-\frac{d-1}{d}\,t}\right)\right).
\end{align*}
Now choose
\[
    t = \frac{d}{d-1}\left(1-
    \frac{\log((j+\frac{d}{d-1})\log n)}{\log n}\right).
\]
So that
\[
    n\binom{n-1}{j}\exp\left(-(n-1-j)
    n^{-\frac{d-1}{d}\,t}\right)
    = O\left(n^{1+j} n^{-j-\frac{d}{d-1}}\right)
    = O(n^{-\frac1{d-1}}),
\]
and
\[
    n^{1-t} = n^{-\frac1{d-1}} \left(j+\tfrac{d}{d-1}\right)
    ^{\frac d{d-1}}(\log n)^{\frac d{d-1}}
    = O\left(n^{-\frac{\varepsilon}{d-1}}
    (\log n)^{\frac{d}{d-1}}\right),
\]
uniformly for $j=O(n^{\frac{1-\varepsilon}{d}})$. Thus
\[
    \mathbb{E}[L_{d,d-1}(n,j)]
    = O\left(n^{-\frac{\varepsilon}{d-1}}
    (\log n)^{d-2+\frac d{d-1}}
    + n^{-\frac1{d-1}}\right) \to 0.
\]
This proves the theorem. \qed

A more precise asymptotic estimate for $\mb{E}[L_{d,d-1}(n,j)]$ will
be derived in Section~\ref{sec:d-large}; see \eqref{Ldd}.
Another easy special case is $k=1$, which is dual to the case $k=d$
because we have
\[
    \mathbb{E}[L_{d,1}(n,j)] = \mathbb{E}[L_{d,d}(n,n-1-j)].
\]
Thus, by (\ref{int-ELdd}), we have
\begin{align*}
    \mathbb{E}[L_{d,1}(n,j)]
    &= \frac{n}{(d-1)!}\binom{n-1}{j}
    \int_0^1 t^{n-1-j} (1-t)^j (-\log t)^{d-1} \text{d}t\\
    &\sim \frac{n^{j+1}}{(d-1)!j!}
    \int_0^\infty e^{-nt} t^{j+d-1} \text{d}t\\
    &\sim \binom{j+d-1}{j} n^{-d+1},
\end{align*}
for large $n$ and $0\le j=o(\sqrt{n})$.

\begin{figure}[!htbp]
\begin{center}
\begin{minipage}{0.4\textwidth}
\begin{tikzpicture} [scale=0.50]
\draw(3,10.00)node{\small $\sum_{0\le j\le m}L_{d,k}(n,j)$};%
\draw(5.80,-1.0)node{\small $m$};%
\draw[line width=1pt](0,0)--(10.20,0);%
\foreach \x/\k in
{0.20/0,2.20/20,4.20/40,6.20/60, 8.20/80,10.20/100} { %
\draw[line width=1pt](\x,-0.2)--(\x,0);%
\draw(\x,-0.5)circle(0pt)node[black]{\tiny$\k$}; }%
\draw[line width=1pt](0,0)--(0,10.00);%
\foreach \y/\k in
{0.00/0,2.00/20,4.00/40,6.00/60, 8.00/80,10.00/100} { %
\draw[line width=1pt](-0.2,\y)--(0,\y);%
\draw(-1,\y)circle(0pt)node[black]{\tiny$\k$}; }%
\draw(7.20,2.38)node{\tiny\color{red}($d=2,k=1$)};%
\draw[line width=1pt,color=red](0.20,0.00) \foreach \x/\y in
{0.40/0.00,0.60/0.01,0.80/0.03,1.00/0.04,1.20/0.06,
1.40/0.09,1.60/0.12,1.80/0.15,2.00/0.19,2.20/0.23,
2.40/0.28,2.60/0.34,2.80/0.40,3.00/0.48,3.20/0.55,
3.40/0.62,3.60/0.69,3.80/0.77,4.00/0.88,4.20/0.98,
4.40/1.10,4.60/1.23,4.80/1.35,5.00/1.47,5.20/1.58,
5.40/1.72,5.60/1.88,5.80/2.05,6.00/2.21,6.20/2.38,
6.40/2.56,6.60/2.76,6.80/2.97,7.00/3.20,7.20/3.43,
7.40/3.70,7.60/4.01,7.80/4.30,8.00/4.63,8.20/4.93,
8.40/5.29,8.60/5.65,8.80/6.05,9.00/6.48,9.20/6.93,
9.40/7.45,9.60/8.05,9.80/8.64,10.00/9.46,10.10/10.00} {-- (\x,\y)} ;
\draw(7.20,6.35)node{\tiny\color{blue}($d=3,k=2$)};%
\draw[line width=1pt,color=blue](0.20,0.03) \foreach \x/\y in
{0.40/0.10,0.60/0.21,0.80/0.34,1.00/0.49,1.20/0.66,
1.40/0.85,1.60/1.04,1.80/1.22,2.00/1.40,2.20/1.60,
2.40/1.81,2.60/2.04,2.80/2.26,3.00/2.47,3.20/2.70,
3.40/2.95,3.60/3.19,3.80/3.43,4.00/3.67,4.20/3.90,
4.40/4.16,4.60/4.41,4.80/4.64,5.00/4.88,5.20/5.11,
5.40/5.36,5.60/5.63,5.80/5.85,6.00/6.08,6.20/6.35,
6.40/6.57,6.60/6.85,6.80/7.08,7.00/7.29,7.20/7.52,
7.40/7.74,7.60/7.95,7.80/8.17,8.00/8.38,8.20/8.59,
8.40/8.79,8.60/8.97,8.80/9.15,9.00/9.33,9.20/9.50,
9.40/9.67,9.60/9.80,9.80/9.90,10.00/9.97,10.10/10.00} {-- (\x,\y)} ;
\draw(5.20,0.78)node{\tiny\color{blue}($d=3,k=1$)};%
\draw[line width=1pt,color=blue](0.20,0.00) \foreach \x/\y in
{0.40/0.00,0.60/0.00,0.80/0.00,1.00/0.00,1.20/0.00,
1.40/0.01,1.60/0.01,1.80/0.01,2.00/0.01,2.20/0.02,
2.40/0.03,2.60/0.04,2.80/0.05,3.00/0.06,3.20/0.07,
3.40/0.08,3.60/0.10,3.80/0.12,4.00/0.14,4.20/0.16,
4.40/0.19,4.60/0.22,4.80/0.27,5.00/0.30,5.20/0.34,
5.40/0.39,5.60/0.45,5.80/0.52,6.00/0.60,6.20/0.68,
6.40/0.77,6.60/0.88,6.80/1.00,7.00/1.15,7.20/1.28,
7.40/1.45,7.60/1.64,7.80/1.81,8.00/2.04,8.20/2.29,
8.40/2.61,8.60/2.97,8.80/3.37,9.00/3.86,9.20/4.43,
9.40/5.14,9.60/5.95,9.80/7.01,10.00/8.57,10.10/10.00} {-- (\x,\y)} ;
\draw(7.20,8.73)node{\tiny\color{black}($d=4,k=3$)};%
\draw[line width=1pt,color=black](0.20,0.16) \foreach \x/\y in
{0.40/0.56,0.60/0.97,0.80/1.35,1.00/1.75,1.20/2.12,
1.40/2.53,1.60/2.92,1.80/3.28,2.00/3.67,2.20/4.03,
2.40/4.35,2.60/4.68,2.80/5.00,3.00/5.29,3.20/5.58,
3.40/5.87,3.60/6.16,3.80/6.42,4.00/6.65,4.20/6.91,
4.40/7.16,4.60/7.38,4.80/7.58,5.00/7.77,5.20/7.97,
5.40/8.14,5.60/8.33,5.80/8.48,6.00/8.61,6.20/8.73,
6.40/8.89,6.60/9.02,6.80/9.13,7.00/9.24,7.20/9.34,
7.40/9.44,7.60/9.52,7.80/9.60,8.00/9.68,8.20/9.74,
8.40/9.79,8.60/9.84,8.80/9.89,9.00/9.92,9.20/9.95,
9.40/9.97,9.60/9.99,9.80/9.99,10.00/10.00,10.10/10.00}{-- (\x,\y)};%
\draw(5.20,3.31)node{\tiny\color{black}($d=4,k=2$)};%
\draw[line width=1pt,color=black](0.20,0.00) \foreach \x/\y in
{0.40/0.00,0.60/0.02,0.80/0.05,1.00/0.08,1.20/0.11,
1.40/0.15,1.60/0.19,1.80/0.24,2.00/0.30,2.20/0.37,
2.40/0.44,2.60/0.51,2.80/0.61,3.00/0.70,3.20/0.81,
3.40/0.90,3.60/1.02,3.80/1.15,4.00/1.28,4.20/1.41,
4.40/1.56,4.60/1.71,4.80/1.87,5.00/2.02,5.20/2.23,
5.40/2.42,5.60/2.65,5.80/2.84,6.00/3.08,6.20/3.31,
6.40/3.57,6.60/3.81,6.80/4.11,7.00/4.39,7.20/4.69,
7.40/4.96,7.60/5.29,7.80/5.62,8.00/5.93,8.20/6.27,
8.40/6.64,8.60/7.03,8.80/7.43,9.00/7.81,9.20/8.21,
9.40/8.61,9.60/9.06,9.80/9.45,10.00/9.84,10.10/10.00} {-- (\x,\y)} ;
\draw(8.20,0.28)node{\tiny\color{black}($d=4,k=1$)};%
\draw[line width=1pt,color=black](0.20,0.00) \foreach \x/\y in
{0.40/0.00,0.60/0.00,0.80/0.00,1.00/0.00,1.20/0.00,
1.40/0.00,1.60/0.00,1.80/0.00,2.00/0.00,2.20/0.00,
2.40/0.00,2.60/0.00,2.80/0.01,3.00/0.01,3.20/0.01,
3.40/0.01,3.60/0.01,3.80/0.02,4.00/0.02,4.20/0.03,
4.40/0.04,4.60/0.05,4.80/0.06,5.00/0.07,5.20/0.08,
5.40/0.10,5.60/0.11,5.80/0.13,6.00/0.15,6.20/0.18,
6.40/0.21,6.60/0.25,6.80/0.29,7.00/0.34,7.20/0.41,
7.40/0.47,7.60/0.55,7.80/0.64,8.00/0.76,8.20/0.88,
8.40/1.06,8.60/1.25,8.80/1.52,9.00/1.85,9.20/2.23,
9.40/2.84,9.60/3.59,9.80/4.85,10.00/7.10,10.10/10.00} {-- (\x,\y)} ;
\end{tikzpicture}
\end{minipage}
\hspace{0.5cm}
\begin{minipage}{0.4\textwidth}
\begin{tikzpicture} [scale=0.50]
\draw(3,10.00)node{\small $\sum_{0\le j\le m}L_{d,k}(n,j)$};%
\draw(5.80,-1)node{\small $m$};%
\draw[line width=1pt](0,0)--(10.20,0);%
\foreach \x/\k in {0.20/0,2.20/1000,4.20/2000,
6.20/3000,8.20/4000,10.20/5000} { %
\draw[line width=1pt](\x,-0.2)--(\x,0);
\draw(\x,-0.5)circle(0pt)node[black]{\tiny$\k$}; }%
\draw[line width=1pt](0,0)--(0,10.00);%
\foreach \y/\k in {0.00/0,2.00/1000,4.00/2000, 6.00/3000,8.00/4000,
10.00/5000} {%
\draw[line width=1pt](-0.2,\y)--(0,\y);%
\draw(-1,\y)circle(0pt)node[black]{\tiny$\k$}; }%
\draw(7.40,2.53)node{\tiny\color{red}($d=2,k=1$)};%
\draw[line width=1pt,color=red](0.20,0.00) \foreach \x/\y in
{0.40/0.00,0.60/0.01,0.80/0.02,1.00/0.03,1.20/0.05,
1.40/0.07,1.60/0.10,1.80/0.13,2.00/0.17,2.20/0.22,
2.40/0.26,2.60/0.31,2.80/0.37,3.00/0.44,3.20/0.50,
3.40/0.58,3.60/0.66,3.80/0.75,4.00/0.84,4.20/0.94,
4.40/1.05,4.60/1.16,4.80/1.28,5.00/1.41,5.40/1.68,
5.60/1.83,5.80/1.99,6.00/2.16,6.20/2.34,6.40/2.53,
6.60/2.73,6.80/2.94,7.00/3.16,7.20/3.39,7.40/3.64,
7.60/3.90,7.80/4.18,8.00/4.48,8.20/4.79,8.40/5.12,
8.60/5.47,8.80/5.85,9.00/6.26,9.20/6.71,9.40/7.18,
9.60/7.72,9.80/8.32,10.00/9.02,10.20/10.00} {-- (\x,\y)} ;
\draw(7.40,6.47)node{\tiny\color{blue}($d=3,k=2$)};%
\draw[line width=1pt,color=blue](0.20,0.00) \foreach \x/\y in
{0.40/0.06,0.60/0.15,0.80/0.27,1.00/0.41,1.20/0.56,
1.40/0.73,1.60/0.91,1.80/1.10,2.00/1.29,2.20/1.49,
2.40/1.70,2.60/1.92,2.80/2.14,3.00/2.36,3.20/2.59,
3.40/2.82,3.60/3.06,3.80/3.29,4.00/3.53,4.20/3.77,
4.40/4.02,4.60/4.27,4.80/4.51,5.00/4.76,5.40/5.25,
5.60/5.50,5.80/5.74,6.00/5.99,6.20/6.23,6.40/6.47,
6.60/6.71,6.80/6.95,7.00/7.18,7.20/7.41,7.40/7.64,
7.60/7.87,7.80/8.09,8.00/8.30,8.20/8.51,8.40/8.71,
8.60/8.91,8.80/9.09,9.00/9.27,9.20/9.43,9.40/9.59,
9.60/9.73,9.80/9.85,10.00/9.95,10.20/10.00} {-- (\x,\y)} ;
\draw(5.40,0.84)node{\tiny\color{blue}($d=3,k=1$)};%
\draw[line width=1pt,color=blue](0.20,0.00) \foreach \x/\y in
{0.40/0.00,0.60/0.00,0.80/0.00,1.00/0.00,1.20/0.00,
1.40/0.00,1.60/0.01,1.80/0.01,2.00/0.01,2.20/0.02,
2.40/0.02,2.60/0.03,2.80/0.04,3.00/0.05,3.20/0.06,
3.40/0.07,3.60/0.09,3.80/0.11,4.00/0.13,4.20/0.15,
4.40/0.18,4.60/0.21,4.80/0.24,5.00/0.28,5.40/0.38,
5.60/0.44,5.80/0.50,6.00/0.57,6.20/0.65,6.40/0.74,
6.60/0.83,6.80/0.95,7.00/1.07,7.20/1.21,7.40/1.36,
7.60/1.53,7.80/1.73,8.00/1.94,8.20/2.19,8.40/2.47,
8.60/2.78,8.80/3.14,9.00/3.55,9.20/4.04,9.40/4.63,
9.60/5.33,9.80/6.25,10.00/7.50,10.20/10.00} {-- (\x,\y)} ;
\draw(7.40,8.83)node{\tiny\color{black}($d=4,k=3$)};%
\draw[line width=1pt,color=black](0.20,0.00) \foreach \x/\y in
{0.40/0.35,0.60/0.76,0.80/1.18,1.00/1.59,1.20/2.00,
1.40/2.40,1.60/2.79,1.80/3.16,2.00/3.53,2.20/3.88,
2.40/4.22,2.60/4.55,2.80/4.87,3.00/5.18,3.20/5.48,
3.40/5.77,3.60/6.04,3.80/6.31,4.00/6.56,4.20/6.80,
4.40/7.03,4.60/7.25,4.80/7.46,5.00/7.67,5.40/8.05,
5.60/8.22,5.80/8.38,6.00/8.54,6.20/8.69,6.40/8.83,
6.60/8.96,6.80/9.08,7.00/9.19,7.20/9.29,7.40/9.39,
7.60/9.48,7.80/9.56,8.00/9.63,8.20/9.70,8.40/9.76,
8.60/9.82,8.80/9.86,9.00/9.90,9.20/9.93,9.40/9.96,
9.60/9.98,9.80/9.99,10.00/10.00,10.20/10.00} {-- (\x,\y)} ;
\draw(5.40,3.45)node{\tiny\color{black}($d=4,k=2$)};%
\draw[line width=1pt,color=black](0.20,0.00) \foreach \x/\y in
{0.40/0.00,0.60/0.01,0.80/0.03,1.00/0.05,1.20/0.07,
1.40/0.10,1.60/0.14,1.80/0.19,2.00/0.24,2.20/0.30,
2.40/0.37,2.60/0.44,2.80/0.52,3.00/0.61,3.20/0.71,
3.40/0.82,3.60/0.93,3.80/1.05,4.00/1.18,4.20/1.32,
4.40/1.47,4.60/1.62,4.80/1.79,5.00/1.96,5.40/2.34,
5.60/2.54,5.80/2.75,6.00/2.97,6.20/3.21,6.40/3.45,
6.60/3.70,6.80/3.96,7.00/4.23,7.20/4.52,7.40/4.82,
7.60/5.13,7.80/5.45,8.00/5.78,8.20/6.12,8.40/6.48,
8.60/6.84,8.80/7.22,9.00/7.60,9.20/8.00,9.40/8.41,
9.60/8.83,9.80/9.25,10.00/9.66,10.20/10.00} {-- (\x,\y)} ;
\draw(8.40,0.27)node{\tiny\color{black}($d=4,k=1$)};
\draw[line width=1pt,color=black](0.20,0.00)%
\foreach \x/\y in
{0.40/0.00,0.60/0.00,0.80/0.00,1.00/0.00,1.20/0.00,
1.40/0.00,1.60/0.00,1.80/0.00,2.00/0.00,2.20/0.00,
2.40/0.00,2.60/0.00,2.80/0.00,3.00/0.00,3.20/0.01,
3.40/0.01,3.60/0.01,3.80/0.01,4.00/0.02,4.20/0.02,
4.40/0.02,4.60/0.03,4.80/0.04,5.00/0.05,5.40/0.07,
5.60/0.08,5.80/0.10,6.00/0.12,6.20/0.14,6.40/0.17,
6.60/0.20,6.80/0.24,7.00/0.29,7.20/0.34,7.40/0.41,
7.60/0.48,7.80/0.57,8.00/0.68,8.20/0.81,8.40/0.96,
8.60/1.15,8.80/1.37,9.00/1.66,9.20/2.02,9.40/2.48,
9.60/3.12,9.80/4.03,10.00/5.52,10.20/10.00} {-- (\x,\y)} ;
\end{tikzpicture}
\end{minipage}
\end{center}
\caption{\emph{Simulated values of $\sum_{0\le j\le m}L_{d,k}(n,j)$
for $n=100$ (left) and $5000$ (right). Interestingly, the simulations
suggest some general pattern that seems independent of the size of the
samples and they are consistent with our analysis since $m$ has to be
very large (compared with $n$).}} \label{fg2}
\end{figure}
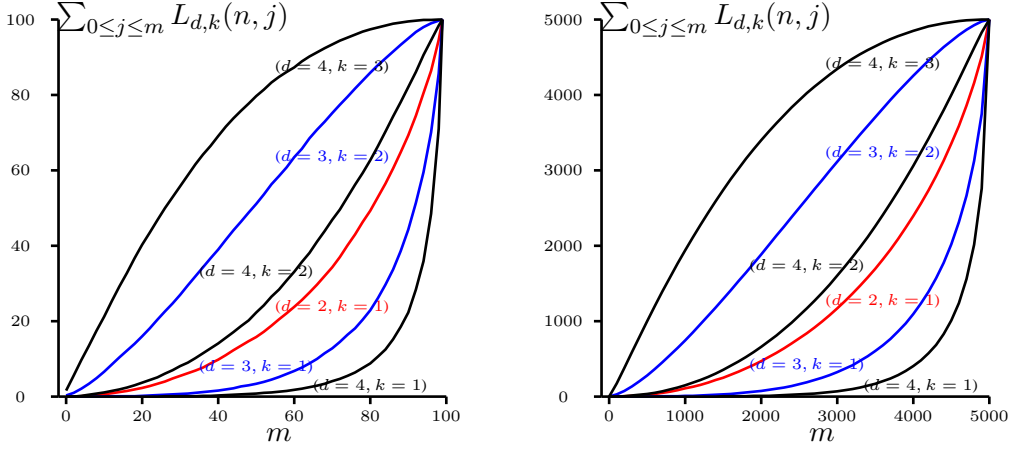

In general, if we are to select the top $K$ representatives using
such clusters of partial dominant skylines, then how large should
$j$ be? That is, what is the minimum $m$ such that $\sum_{0\le j\le
m}L_{d,k}(n,j)>K$? Some simulation results are given in
Figure~\ref{fg2}.

\section{Random samples from simplices}
\label{sec:simplex}

We show in this section that the asymptotic vanishing property of
$k$-dominant skylines occurs not only in the case of the
$d$-dimensional hypercube distribution, but also in the
$d$-dimensional simplex distribution
\[
    S_d=\left\{\v{x}:-1\le x_j\le 0 \text{ and }
    \left\| \v{x}\right\|
    :=\sum_{1\le j\le d}\left|x_j\right| \le 1\right\}.
\]
In particular, $S_2$ is the right triangle
\raisebox{-0.05cm}{\begin{tikzpicture}[scale=0.5]
\draw[line width=0.5pt] (0,0) -- (-0.5,0) -- (0,-0.5) -- (0,0);%
\end{tikzpicture}}. Such a shape implies a negative dependence of the
two coordinates and thus a larger number of skyline points.

Let $M^{[s]}_k(n)$ denote the cardinality of the $k$-dominant
skyline of the set $\mathscr{D} := \{\v{p}_1,\ldots ,\v{p}_n\}$,
where these $n$ points are uniformly and independently distributed
over $S_{d}$. For a point $\v{p}\in S_d$, denote by
$B^{[s]}_k(\v{\v{p}})$ the region of points in $S_d$ that
$k$-dominate $\v{p}$.

\begin{thm}[Asymptotic vanishing property for finite-dimensional
simplex] For $1\le k\le d-1$,
\[
    \mathbb{E}[M^{[s]}_{d,k}(n)]
    \to \left\{\begin{array}{ll}
        0, & \text{if }1\le k\le d-1;\\
        \infty,& \text{if }k=d,
    \end{array}
    \right.
\]
as $n\to\infty$.
\end{thm}
\pf  For $k=d$, it is known (see \cite{CHT11}) that
\begin{align*}
    \mathbb{E}[M^{[s]}_{d,d}(n)]
    &= d! n \int_D\left(1-\left(1-
    {\textstyle\sum}_{1\le i\le d} x_i\right)^d\right)^{n-1}
    \text{d} \v{x} \\
    &=n\sum_{0\le j<d} \binom{d-1}{j}
    (-1)^j\frac{\Gamma(n)\Gamma\left(\frac{j+1}d\right)}
    {\Gamma\left(n+\frac{j+1}{d}\right)}\\
    &= \Gamma\left(\tfrac1d\right) n^{1-\frac1d}\left(
    1+O\left(d n^{-\frac1d}\right)\right),
\end{align*}
where $\Gamma$ denotes the Gamma function. Thus the expected number
of skylines tends to infinity as $n$ goes unbounded.

Consider now $1\le k<d$. It suffices to examine the case $k=d-1$.
For a point $\v{x}\in S_{d}$ ($\v{x}\neq \v{0}$), let $\bm{\xi}:=
\frac{\v{x}}{\|\v{x}\|}$. Then $B_{d-1}^{[s]}(\bm{\xi}) \subset
B^{[s]}_{d-1}(\v{x})$. We now prove that
\begin{equation}
    \left|B^{[s]}_{d-1}(\bm{\xi})\right|
    \ge \frac{1}{d!d^{d}}\qquad( \bm{\xi}\in S_{d},
    \left\| \bm{\xi}\right\| =1). \label{t2}
\end{equation}
Since $\left\| \bm{\xi}\right\|=1$, there is at least one
coordinate $|\xi_j|\ge \frac1d$. Without loss of generality, assume
$|\xi_d| \ge \frac1d$. Then $\sum_{1\le j<d} | \xi_j| \le
\frac{d-1}d$. Let
\[
    T := \{\v{y}\in S_{d}:y_j\le \xi_j
    \text{ for }1\le j\le d-1\text{ and }y_d\le 0\}.
\]
We have $T\subset B^{[s]}_{d-1}(\bm{\xi})$ and
\[
    |T| = |S_d||\xi_d| \ge \frac{1}{d!d^{d}},
\]
since $T$ is itself a simplex. Thus \eqref{t2} holds and we have
\begin{align*}
    \mathbb{E}[M^{[s]}_{d,d-1}(n)]
    &=nd!\int_{S_{d}}\left( 1-d!\left| B_{d-1}^{[s]}
    (\v{x})\right| \right)^{n-1}\mathrm{d}\v{x}\\
    &=O\left( n\left(1-d^{-d}\right)^{n}\right)\\
    &\rightarrow 0,
\end{align*}
as $n\rightarrow \infty $. \qed

We see in such a simplex model that the expected number of
$k$-dominant tends to zero at an \emph{exponential} rate (in $n$),
in contrast to the \emph{polynomial} rate in the hypercube model.
Does the expected number of $k$-dominant skyline points always tend
to zero? Here is a simple, artificial counterexample.

\quad \newline
\textbf{Example 1.} Assume $d=4,k=3$. Let
\[
    A:=\left\{ (-t,-2t,3t,4t):1\le t\le 2\right\} .
\]
Then any two points in $A$ are incomparable (none dominating the
other) by the relation of $k$-dominance. Thus, the number of
$k$-dominant skyline points is equal to $n$ almost surely if
$\v{p}_1,\ldots ,\v{p}_n$ are uniformly and independently
distributed in $A$.

\section{A categorical model}
\label{sec:cm}

The preceding negative results are based on assuming that the points
are generated from some \emph{continuous models}, which are often a good
approximation to situations where the input can assume a sufficiently large
range of different values. What if we assume instead that the inputs are
sampled from some \emph{discrete space}, which is also often encountered
in practical applications? We show in this section that \emph{the 
expected number of $k$-dominant skylines is always linear for 
$1\le k\le d$}, in contrast to the asymptotic zero-infinity property 
we derived above.

Assume that $n$ points $\mathscr{D} := \{\v{p}_1,\dots, \v{p}_n\}$
are chosen uniformly and independently from the product space
\[
    \mathscr{P} := \bigotimes_{1\le j\le d}S_j,
\]
where
\[
    S_j =\{1,2,\ldots ,u_j\}\qquad(u_j\ge2).
\]
Let $M_{d,k}^{[c]}(n)$ denote the number of $k$-dominant skylines in
$\mathscr{D}$. Unlike the continuous cases, the variation of the
random variables $M_{d,k}^{[c]}(n)$ is easier to predict as the
number of possible points in $\mathscr{P}$ is finite. Interestingly,
the first-order asymptotic estimate for the expected value of
$M_{d,k}^{[c]}(n)$ is independent of $k$ for $1\le k\le d$, where
the case $k=d$ gives the expected skyline count.
\begin{thm}[Asymptotic linearity for finite-dimensional categorical
model] The expected number of $k$-dominant skylines satisfies
\begin{align} \label{Mkn-cm}
    \frac{\mb{E}[M_{d,k}^{[c]}(n)]}n\rightarrow \frac 1u
    \qquad(1\le k\le d; d\ge 2),
\end{align}
as $n\rightarrow \infty$, where
\[
    u := \prod_{1\le j\le d}u_j.
\]
\end{thm}
Now the problem is again the excessive number of skyline points.
Such a discrete model exhibits another interesting phenomenon, not
present for continuous model, namely, for fixed $n$, the expected
number of $k$-dominant skyline points is not monotonically
increasing as $d$ grows.

\pf  
Let $\v{x}=(x_1,x_2,\ldots, x_d)\in \mathscr{P}$. Denote by
$B^{[c]}_k(\v{x})$ the set of points in $\mathscr{P}$ that
$k$-dominate $\v{x}$. Then
\begin{align} \label{EMkn-cm}
    \mb{E}[M_{d,k}^{[c]}(n)]
    &= n\mb{P}(\v{p}_1 \text{ is a $k$-dominant
    skyline point})\nonumber \\
    &=\frac{n}{u}\sum_{\v{x}\in
    \mathscr{P}}\left( 1-
    \frac{\left| B_{k}^{[c]}(\v{x})\right|}{u}\right)^{n-1}.
\end{align}
If $\v{y}\in B_k^{[c]}(\v{x})$, then $\v{y}$ is better than or equal
to $\v{x}$ in all coordinates (at least one better) except for the
coordinates, say $j_1,\ldots ,j_{\ell }$ for $0\le \ell \le d-k$.
Thus
\[
    \left| B_d^{[c]}(\v{x})\right|
    = \prod_{1\le j\le d} x_j-1 ,
\]
and for $1\le k<d$
\begin{align}
    \left| B_k^{[c]}(\v{x})\right|
    =\sum_{0\le \ell\le d-k}\sum_{1\le
    j_1<j_2<\cdots <j_\ell \le d}\left(
    \frac{\prod_{1\le i\le d}x_{i}}{
    \prod_{1\le i\le \ell} x_{j_i}}-1\right)
    \prod_{1\le i\le \ell}
    \left(u_{j_i}-x_{j_i}\right) . \label{Bkvx}
\end{align}
Here the product
\[
    \frac{\prod_{1\le i\le d}x_{i}}{
    \prod_{1\le i\le \ell} x_{j_i}}
    = \prod_{i\neq j_r; r=1,\dots,\ell} x_i,
\]
enumerates all possible locations in the $d-\ell$ ($\ge k$)
coordinates that $k$-dominant skyline point can assume, and the
factor ``$-1$" removes the possibility that all $d-\ell$ coordinates
are equal to the corresponding $x_i$. The last product in
\eqref{Bkvx} describes all possible locations for the other $\ell$
coordinates.

Since there is a unique point $\v{1} := (\overbrace{1,\ldots ,1}^d)$
in $\mathscr{P}$ with $\left| B_{k}^{[c]}(\v{1})\right| =0$, all
other terms in the sum on the right-hand side of (\ref{EMkn-cm})
being exponentially small, we obtain \eqref{Mkn-cm}. 
\qed

\medskip
\begin{figure}[tbp]
\begin{center}
\begin{minipage}{0.4\textwidth}
\centering
\begin{tikzpicture} [scale=0.50]
\draw(-1.00,11.00)node{\normalsize Mean};
\draw(5.80,-1.50)node{\normalsize$n$}; \draw[line
width=1pt](0,0)--(10.20,0); \foreach \x/\k in
{0.20/1,1.87/5,3.95/10,6.03/15,8.12/20,10.20/25} { \draw[line
width=1pt](\x,-0.2)--(\x,0);
\draw(\x,-0.5)circle(0pt)node[black]{\normalsize$\k$}; } \draw[line
width=1pt](0,0)--(0,10.00); \foreach \y/\k in
{0.00/1.0,3.33/1.5,6.67/2.0,10.00/2.5} { \draw[line
width=1pt](-0.2,\y)--(0,\y);
\draw(-1,\y)circle(0pt)node[black]{\normalsize$\k$}; } \foreach \y
in {0.67,1.33,2.00,2.67,4.00,4.67,5.33,6.00,7.33,8.00,8.67,9.33}
\draw[line width=1pt](-0.1,\y)--(0,\y); \draw[line
width=1pt,color=red](0.20,0.00) \foreach \x/\y in
{0.62/3.49,1.03/5.69,1.45/7.12,1.87/8.06,2.28/8.67,
2.70/9.05,3.12/9.27,3.53/9.37,3.95/9.39,4.37/9.34,
4.78/9.24,5.20/9.10,5.62/8.94,6.03/8.75,6.45/8.56,
6.87/8.35,7.28/8.13,7.70/7.91,8.12/7.69,8.53/7.47,
8.95/7.25,9.37/7.03,9.78/6.81,10.20/6.59}
{-- (\x,\y)} ;
\end{tikzpicture}
\end{minipage}
\hspace{0.5cm}
\begin{minipage}{0.4\textwidth}
\begin{tikzpicture} [scale=0.50]
\draw(-1.00,11.00)node{\normalsize Mean};
\draw(5.80,-1.50)node{\normalsize$n$}; \draw[line
width=1pt](0,0)--(10.20,0); \foreach \x/\k in
{0.20/25,2.70/250,5.20/500,7.70/750,10.20/1000} { \draw[line
width=1pt](\x,-0.2)--(\x,0);
\draw(\x,-0.5)circle(0pt)node[black]{\normalsize$\k$}; } \draw[line
width=1pt](0,0)--(0,10.00); \foreach \y/\k in
{0.00/0.0,2.50/0.5,5.00/1.0,7.50/1.5,10.00/2.0} { \draw[line
width=1pt](-0.2,\y)--(0,\y);
\draw(-1,\y)circle(0pt)node[black]{\normalsize$\k$}; }%
\foreach \y in {0.50,1.00,1.50,2.00,3.00,3.50,4.00,4.50,5.50,
6.00,6.50,7.00,8.00,8.50,9.00,9.50} \draw[line
width=1pt](-0.1,\y)--(0,\y); \draw(8.89,3.80)node{}; \draw[line
width=1pt,color=red](0.20,9.94) \foreach \x/\y in
{0.46/6.84,0.71/5.13,0.97/4.12,1.23/3.49,1.48/3.06,
1.74/2.78,1.99/2.58,2.25/2.44,2.51/2.35,2.76/2.30,
3.02/2.27,3.28/2.26,3.53/2.27,3.79/2.30,4.05/2.34,
4.30/2.40,4.56/2.47,4.82/2.54,5.07/2.62,5.33/2.71,
5.58/2.81,5.84/2.91,6.10/3.01,6.35/3.12,6.61/3.23,
6.87/3.34,7.12/3.45,7.38/3.57,7.64/3.69,7.89/3.80,
8.15/3.92,8.41/4.04,8.66/4.16,8.92/4.28,9.17/4.40,
9.43/4.52,9.69/4.64,9.94/4.76,10.20/4.89}
{-- (\x,\y)} ;
\end{tikzpicture}
\end{minipage}
\end{center}
\caption{\emph{A graphical rendering of $\mb{E}[M_{d,k}^{[c]}(n)]$
in the discrete space $ \{0,1\}^{d} $ for $d=10$, $k=9$ and
$n=1,\ldots,25$ (left) and $ n=25,\ldots,1000$ (right). }}
\end{figure}
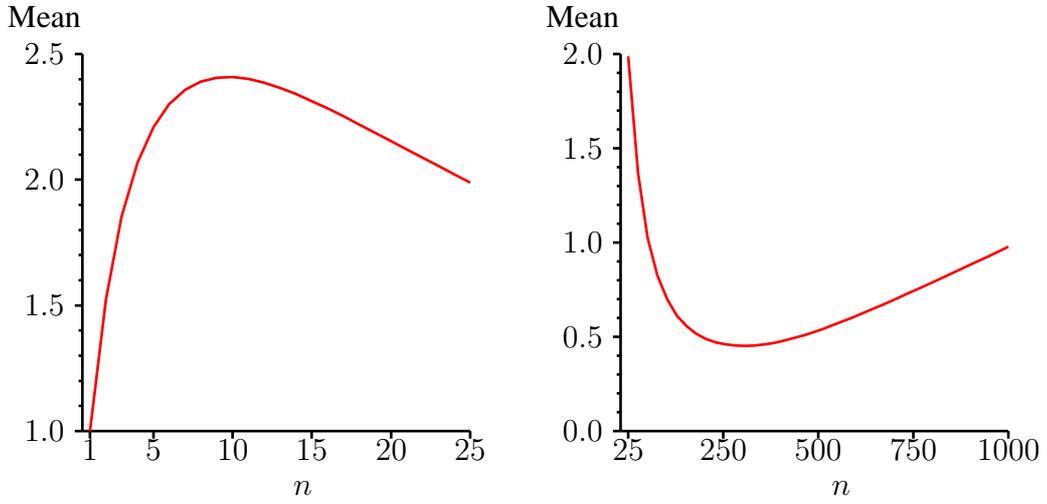

In the special case when all $u_j=2$ for $1\le j\le d$, then
\[
    \left| B_{k}^{[c]}(\v{x})\right|
    =\left( 2^\ell-1\right) \sum_{0\le j \le d-k}
    \binom{d-\ell}{j},
\]
where $\v{x}\in \{1,2\}^d$ and $\ell$ denotes the number of times
``$2$" occurs in $\v{x}$ (and ``$1$" occurring $d-\ell$ times). The
closed-form expression \eqref{EMkn-cm} simplifies
\[
    \mb{E}[M_{d,k}^{[c]}(n)]=\frac{n}{2^{d}}
    \sum_{0\le\ell \le d}\binom{d}{\ell}\left( 1-\frac{
    2^\ell-1}{2^d}\sum_{0\le j\le d-k}
    \binom{d-\ell}{j}\right)^{n-1},
\]
from which it follows that
\[
    \frac{\mb{E}[M_{d,k}^{[c]}(n)]}{n}\rightarrow
    \frac{1}{2^{d}}\quad \text{as }
    n\rightarrow \infty .
\]

\begin{figure}[!htbp]
\begin{center}
\begin{tikzpicture} [scale=0.45]
\draw(-1.00,11.00)node{\normalsize };
\draw(5.0,-1.50)node{\normalsize$n$};%
\draw[line width=1pt](0,0)--(10.10,0);
\draw(0,-0.5)node[black]{\normalsize$1$};%
\foreach \x/\k in {1.25/1,2.50/2,3.75/3,
5.00/4,6.25/5,7.50/6,8.75/7,10.00/8} {%
\draw[line width=0.9pt](\x,-0.2)--(\x,0);
\draw(\x,-0.5)circle(0pt)node[black]{\normalsize$2^\k$}; }
\draw[line width=1pt](0,0)--(0,10.10); \foreach \y/\k in
{0.00/0.0,5.00/0.5,10.00/1.0} { %
\draw[line width=0.9pt](-0.2,\y)--(0,\y);
\draw(-1,\y)circle(0pt)node[black]{\normalsize$\k$}; }%
\draw[line width=1pt,color=red](0.00,10.00)%
\foreach \x/\y in {1.25/2.91,2.51/0.89,3.76/0.40,5.02/0.32,
6.27/0.31,7.53/0.31,8.78/0.31,10.04/0.31} {-- (\x,\y)} ;
\draw(1.5,1)node{\normalsize\color{red}$k=3$}; %
\foreach \x/\y in {0.00/10.00,1.25/2.91,2.51/0.89,
3.76/0.40,5.02/0.32,6.27/0.31,7.53/0.31,8.78/0.31,10.04/0.31} {
\draw(\x,\y)circle(2pt); } %
\draw[line width=1pt,color=blue](0.00,10.00) %
\foreach \x/\y in {1.25/4.77,2.51/2.14,3.76/0.91,5.02/0.44,
6.27/0.32,7.53/0.31,8.78/0.31,10.04/0.31} {-- (\x,\y)} ;
\draw(4,2)node{\normalsize\color{blue}$k=4$}; %
\foreach \x/\y in {0.00/10.00,1.25/4.77,2.51/2.14,3.76/0.91,
5.02/0.44,6.27/0.32,7.53/0.31,8.78/0.31,10.04/0.31} {
\draw(\x,\y)+(18:2pt)-- +(90:2pt)-- +(162:2pt)-- +(234:2pt)--
+(306:2pt)-- cycle; } %
\draw[line width=1pt,color=red](0.00,10.00) %
\foreach \x/\y in {1.25/7.94,2.51/5.78,3.76/3.71,5.02/2.07,
6.27/1.05,7.53/0.53,8.78/0.34,10.04/0.31} {-- (\x,\y)} ;
\draw(6,3)node{\normalsize\color{red}$k=5$}; %
\foreach \x/\y in {0.00/10.00,1.25/7.94,2.51/5.78,3.76/3.71,
5.02/2.07,6.27/1.05,7.53/0.53,8.78/0.34,10.04/0.31} {
\draw(\x,\y)+(0:2pt)-- +(90:2pt)-- +(180:2pt)-- cycle; }
\end{tikzpicture}
\quad
\begin{tikzpicture} [scale=0.45]
\draw(-1.00,11.00)node{\normalsize };
\draw(5.0,-1.50)node{\normalsize$n$};
\draw[line width=1pt](0,0)--(10.10,0);
\draw(0,-0.5)node[black]{\normalsize$1$};
\foreach \x/\k in {1.67/1,3.33/2,5.00/3,6.67/4,8.33/5,10.00/6}{%
\draw[line width=0.9pt](\x,-0.2)--(\x,0);
\draw(\x,-0.5)circle(0pt)node[black]{\normalsize$5^\k$};} %
\draw[line width=1pt](0,0)--(0,10.10); \foreach \y/\k in
{0.00/0.0,5.00/0.5,10.00/1.0} { %
\draw[line width=0.9pt](-0.2,\y)--(0,\y);
\draw(-1,\y)circle(0pt)node[black]{\normalsize$\k$}; } %
\draw[line width=1pt,color=red](0.00,10.00) %
\foreach \x/\y in {1.67/0.63,3.33/0.02,
5.00/0.00,6.67/0.00,8.33/0.00,10.00/0.00} {-- (\x,\y)};%
\draw(0.6,1)node{\normalsize\color{red}$k=3$}; %
\foreach \x/\y in {0.00/10.00,1.67/0.63,3.33/0.02,5.00/0.00,
6.67/0.00,8.33/0.00,10.00/0.00} { %
\draw(\x,\y)circle(2pt); }%
\draw[line width=1pt,color=blue](0.00,10.00)%
\foreach \x/\y in {1.67/3.19,3.33/0.45,
5.00/0.03,6.67/0.00,8.33/0.00,10.00/0.00} {-- (\x,\y)} ;%
\draw(3.60,2.00)node{\normalsize\color{blue}$k=4$};%
\foreach \x/\y in {0.00/10.00,1.67/3.19,3.33/0.45,5.00/0.03,
6.67/0.00,8.33/0.00,10.00/0.00} { %
\draw(\x,\y)+(18:2pt)-- +(90:2pt)-- +(162:2pt)-- +(234:2pt)--
+(306:2pt)-- cycle; }%
\draw[line width=1pt,color=red](0.00,10.00) %
\foreach \x/\y in {1.67/7.68,3.33/3.99,
5.00/1.15,6.67/0.16,8.33/0.01,10.00/0.00} {-- (\x,\y)};%
\draw(5.80,3.00)node{\normalsize\color{red}$k=5$}; %
\foreach \x/\y in {0.00/10.00,1.67/7.68,3.33/3.99,5.00/1.15,
6.67/0.16,8.33/0.01,10.00/0.00} { \draw(\x,\y)+(0:2pt)-- +(90:2pt)--
+(180:2pt)-- cycle; }
\end{tikzpicture}
\caption{\emph{Two plots of the ratio $\mb{E}[M_{d,k}^{[c]}(n)]/n$
when $d=5$, $k=3,4,5$ (here the case $k=5$ corresponds to the
skyline), $u_{i}\equiv2$ (left) and $u_{i}\equiv5$ (right). All
curves in the left figure tend to the limit $2^{-5}=0.03125$ while
those in the right to $5^{-5}=0.00032$, which is almost zero.}}
\end{center}
\end{figure}
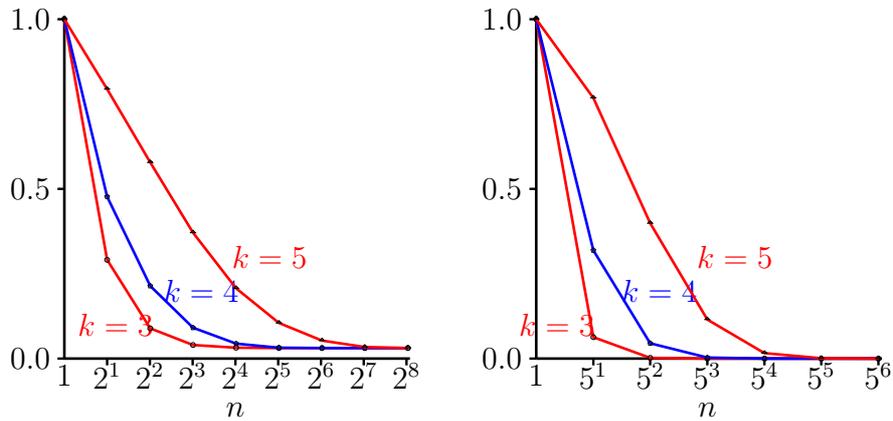

Since the product space $\mathscr{P}$ is finite, we can indeed fully
characterize the asymptotic distribution of $M_{d,k}^{[c]}(n)$.
\begin{thm}[Asymptotic binomial distribution for finite-dimensional
categorical model] The distribution of $M_{d,k}^{[c]}(n)$ is
asymptotically equivalent to a binomial distribution with parameters
$n$ and $1/u$.
\end{thm}
\pf  
Let $X_n$ denote the number of $j$'s for which $\v{p}_j
=(1,\dots,1)$, $1\le j\le n$. Then, obviously, $X_n$ is binomially
distributed with parameters $n$ and $1/u$, namely,
\[
    \mathbb{P}(X_n=\ell) =
    \binom{n}{\ell}\frac1{u^\ell}
    \left(1-\frac1u\right)^{n-\ell} \qquad
    (0\le \ell \le n).
\]
Now if one of the points $\v{p}_j$ equals $(1,\ldots ,1)$, then
$M_{d,k}^{[c]}(n)=X_n$. Thus
\[
    \mb{P}\left( M_{d,k}^{[c]}(n)\neq X_n\right)
    \leq \mb{P}\left( \v{p}_j\neq (1,\ldots ,1)\right)
    =\left( 1-\frac1u\right) ^{n}\rightarrow 0,
\]
and thus the distribution of $M_{d,k}^{[c]}(n)$ is asymptotic to the
distribution of $X_n$. \qed

In particular, we see that the variance of $M_{d,k}^{[c]}(n)$ is
also asymptotically linear
\[
    \frac{\mb{V}[M_{d,k}^{[c]}(n)]}{n}
    \to \frac1{u}\left(1-\frac1u\right)
    \qquad(1\le k\le d).
\]

The consideration can be easily extended to the case of non-uniform
discrete distributions. More generally, assume that the data set is
sampled from the set $\{\v{a} _{1},\ldots ,\v{a}_{m}\}\subset
\mathscr{P}$ and each point is endowed with the probability
$\mb{P}(\v{a}_j)$. Let $p_k(\v{a}_j)$ be the probability that
$\v{a}_j$ is $k$-dominated, that is, $p_k(\v{a}_j)$ is equal to the
sum of $\mb{P}(\v{a}_i)$ such that $\v{a}_i$ $k$-dominates
$\v{a}_j$. Then the expected number of $k$-dominant skyline points
satisfies
\[
    \mb{E}[M_{d,k}^{[c]}(n)]
    =n\sum_{1\le j\le m}\mb{P}(\v{a}_j)\left( 1-p_k(
    \v{a}_j)\right)^{n-1}.
\]
Let
\[
    q_k := \sum_{\substack{p_k(\v{a}_j)=0\\
    1\le j\le m}} \mb{P}(\v{a}_j)
\]
be the probability of points in $\{\v{a}_1,\ldots ,\v{a}_m\}$ that
are not $k$-dominated. Then since the expected number of
$k$-dominant is expressed as a finite sum, we have
\[
    \frac{\mb{E}[M_{d,k}^{[c]}(n)]}{n}\rightarrow q_{k},
    \quad \text{as }n\rightarrow
    \infty .
\]
Note that $p_{k}$ may range from zero to one.

\section{Uniform asymptotic estimates for $\mb{E}[M_{d,d-1}(n)]$}
\label{sec:d-large}

We derive in this section two uniform asymptotic estimates for
$\mb{E}[M_{d,d-1}(n)]$ in two overlapping ranges. To state our
results, we need to introduce the Lambert $W$-function (see
\cite{CGHK96}), which is implicitly defined by the equation
\begin{align} \label{lambert-w}
    W(z)e^{W(z)} =z .
\end{align}
For our purpose, we take $W$ to be the principal branch that is
positive for positive $z$ and satisfies the asymptotic approximation
\begin{align} \label{Wx}
    W(x) = \log x - \log\log x + \frac{\log\log x}{\log x}
    + O\left(\frac{(\log\log x)^2}{(\log x)^2} \right),
\end{align}
for large $x$.

Our first asymptotic estimate covers $d$ in the range
\[
    3\le d\le \sqrt{\frac{2\log n}{W(2\log n)+K}},
\]
where $K\to\infty$ with $n$, and the second the range
\[
    (\log n)^{1/3}\ll
    d\le 2\sqrt{\frac{\log n}{W(\log n)-C}},
\]
for some constant $C>0$. The upper bounds of the two ranges do not
differ significantly but are sufficient for our purposes of proving
the threshold phenomenon, which we discuss in the next section.

Very roughly, the expected number of $(d-1)$-dominant skylines is
asymptotically negligible in the first range, and undergoes the
phase transition from being almost zero to unbounded in the
second.

\begin{thm}[Uniform estimate for large $n$ and moderate $d$]
\label{thm:ud} If $d\ge3$ and
\begin{align} \label{d-rg1}
    \frac{2\log n}{d^2}-W(2\log n)\to\infty,
\end{align}
then
\begin{align} \label{dm1-dom}
    \mathbb{E}[M_{d,d-1}(n)] =
    \frac{n^{-\frac1{d-1}}}{d-1}\,\Gamma\left(\frac1{d-1}\right)^d
    \left(1+O\left(dn^{-\frac1{(d-1)(d-2)}}\right)\right),
\end{align}
uniformly in $d$ for large $n$.
\end{thm}
Note that if $d$ is of the form
\[
    d = \left\lfloor \sqrt{\frac{2\log n}
    {W(2\log n)+2v}}\right\rfloor,
\]
then
\[
    d n^{-\frac1{(d-1)(d-2)}} = e^{-v}\left(1+O\left(
    \frac{(1+|v|)W(2\log n)^{3/2}}{\sqrt{\log n}}\right)\right),
\]
which becomes $o(1)$ if $v\to\infty$.

On the other hand, when $d=2$, we have, by \eqref{EM},
\[
    \mathbb{E}[M_{d,d-1}(n)]
    = n\int_0^1\!\!\int_0^1\left(1-x-y+xy\right)^{n-1}
    \text{d} x\text{d} y
    = \frac1n.
\]

\medskip

\pf  
We again begin with the integral representation \eqref{EM},
where $B_{d-1}(\v{x})$ is given in \eqref{Bd}.

By the elementary inequalities (see \cite{BHLT01})
\[
    e^{-nt}(1-nt^2)\le (1-t)^n \le e^{-nt}
    \qquad(n\ge1; t\in[0,1]),
\]
we have
\[
    E_{n,d}-E_{n,d}' \le \mathbb{E}[M_{d,d-1}(n+1)] \le E_{n,d},
\]
where
\begin{align*}
    E_{n,d}  &:= n\int_{[0,1]^d}
    e^{-n|B_{d-1}(\v{x})|}\text{d} \v{x},\\
    E_{n,d}' &:= n^2 \int_{[0,1]^d} |B_{d-1}(\v{x})|^2
    e^{-n|B_{d-1}(\v{x})|}\text{d} \v{x}.
\end{align*}
We will see that $E_{n,d}'$ is asymptotically of smaller order
than $E_{n,d}$. The intuition here is that most contribution to the
integral comes from $\v{x}$ for which $|B_{d-1}(\v{x})|$ is small,
implying that $(1-|B_{d-1}(\v{x})|)^n$ is close to
$e^{-n|B_{d-1}(\v{x})|}$. Also replacing $n+1$ by $n$ in the
resulting asymptotic approximation gives rise only to smaller order
errors. However, the uniform error bound represents the most delicate 
part of our proof.

We start with the asymptotic evaluation of $E_{n,d}$. By making the
change of variables $x_j \mapsto \frac{y_j}N$, where $N :=
n^{\frac1{d-1}}$,
\begin{align}
    E_{n,d} &= N^{-1} \int_{[0,N]^d}
    e^{-y_1\cdots y_d \left(\frac1{y_1}
    +\cdots+\frac1{y_d}\right) +
    \frac{d-1}{N}y_1\cdots y_d}\text{d} \v{y}\nonumber \\
    &= N^{-1} \left(\phi_d(n)-f_d(n)
    +R_d(n) \right),\label{End-Rnd}
\end{align}
where
\begin{align*}
    \phi_d(n) & := \int_{\mathbb{R}_+^d}
    e^{-y_1\cdots y_d \left(\frac1{y_1}
    +\cdots+\frac1{y_d}\right)}\text{d} \v{y},\\
    f_d(n) &:= \left(\int_{\mathbb{R}_+^d}-
    \int_{[0,N]^d}\right)
    e^{-y_1\cdots y_d \left(\frac1{y_1}
    +\cdots+\frac1{y_d}\right)}\text{d} \v{y},\\
    R_d(n) &:= \int_{[0,N]^d}
    e^{-y_1\cdots y_d \left(\frac1{y_1}
    +\cdots+\frac1{y_d}\right)}\left(
    e^{\frac{d-1}{N}y_1\cdots y_d}-1\right)\text{d} \v{y}.
\end{align*}
We focus on the evaluation of the integral $\phi_d(n)$, leaving the
lengthier estimation of the two error terms $f_d(n)$ and $R_d(n)$ to
Appendix A.

We now carry out the change of variables $t_j := \prod_{\ell\not=j}
y_\ell$ for $1\le j\le d$, the Jacobian being
\[
    \frac{\partial(y_1,\dots,y_d)}{\partial(t_1,\cdots,t_d)}
    :=\left[\begin{array}{ccc}
        \frac{\partial y_1}{\partial t_1} & \cdots
        & \frac{\partial y_1}{\partial t_d} \\
        \vdots & \ddots & \vdots \\
        \frac{\partial y_d}{\partial t_1} & \cdots
        & \frac{\partial y_d}{\partial t_d},
    \end{array}\right]
\]
whose determinant is equal to $1/\det J$, where
\[
    J :=\frac{\partial(t_1,\dots,t_d)}
    {\partial(y_1,\cdots,y_d)}.
\]
Note that the entries of $J$ satisfy
\[
    J_{i,j}=\left\{\begin{array}{ll}
        0,& \text{if } i=j;\\
        \displaystyle \frac{y_1\cdots y_d}{y_iy_j},
        & \text{if }i\neq j.
    \end{array}\right.
\]
It follows that
\[
    \det J = (y_1\cdots y_d)^{d-2} \det T,
\]
where $T$ is a $d\times d$ matrix with $T_{i,i}=0$ and $T_{i,j}=1$
for $i\neq j$. The determinant of $T$ is seen to be $(-1)^{d-1}
(d-1)$ by adding all rows of $T$ to the first, by taking the factor
$d-1$ out, and then by subtracting the first row from all other
rows. Thus we have
\begin{align*}
    \det J &= (-1)^{d-1}(d-1)(y_1\cdots y_d)^{d-2}\\
    &= (-1)^{d-1}(d-1)(t_1\cdots t_d)^{\frac{d-2}{d-1}}.
\end{align*}
Thus, by the integral representation of the Gamma function
\[
    \Gamma(x) = \int_0^\infty t^{x-1} e^{-t} \text{d} t
    \qquad(x>0),
\]
we obtain
\begin{align*}
    \phi_d(n) &= \frac{1}{d-1}
    \int_{\mathbb{R}_+^d} e^{-(t_1+\cdots +t_d)}
    (t_1\cdots t_d)^{-\frac{d-2}{d-1}}\text{d} \v{t}\\
    &=\frac{1}{d-1}\left(\int_0^\infty
    e^{-u} u^{-\frac{d-2}{d-1}}\text{d} u\right)^d\\
    &= \frac1{d-1}\,\Gamma\left(\frac1{d-1}\right)^d.
\end{align*}
We will prove in Appendix A that
\begin{align}
    \frac{f_d(n)}{\phi_d(n)}
    &= O\left(d n^{-\frac1{(d-1)(d-2)}} \right),\nonumber \\
    \frac{R_d(n)}{\phi_d(n)} &= O\left(d 2^{-d}
    n^{-\frac1{d-1}} \right).\label{Rdn-ratio}
\end{align}

In a similar manner, we have
\begin{align*}
    E_{n,d}' &= O\left(n^2\int_{\mathbb{R}_+^d}
    \left(x_1\cdots x_d{\textstyle \sum}_{1\le j\le d}
    \tfrac1{x_j}\right)^2 e^{-n x_1\cdots x_d
    \sum_{1\le j\le d}\tfrac1{x_j}} \text{d}\v{x}\right)\\
    &= O\left(\frac{n^{-\frac2{d-1}}}{d-1}\int_{\mathbb{R}_+^d}
    \left(t_1+\cdots+t_d\right)^2 e^{-(t_1+\cdots+t_d)}
    (t_1\cdots t_d)^{-\frac{d-2}{d-1}}\text{d}\v{t}\right).
\end{align*}
The last integral in a more general form can be evaluated as
follows. Let $[z^n]f(z)$ denote the coefficient of $z^n$ in the
Taylor expansion of $f$.
\begin{align*}
    &\int_{\mathbb{R}_+^d}
    \left(t_1+\cdots+t_d\right)^j e^{-(t_1+\cdots+t_d)}
    (t_1\cdots t_d)^{-\frac{d-2}{d-1}}\text{d}\v{t}\\
    &\qquad= j![z^j] \int_{\mathbb{R}_+^d}
    e^{-(1-z)(t_1+\cdots+t_d)}
    (t_1\cdots t_d)^{-\frac{d-2}{d-1}}\text{d}\v{t}\\
    &\qquad= j![z^j] \frac{\Gamma(\frac1{d-1})^d}
    {(1-z)^{\frac d{d-1}}}\\
    &\qquad= j!\Gamma\left(\frac1{d-1}\right)^d
    \binom{\frac1{d-1}+j}{j},
\end{align*}
for $j\ge0$. Thus
\begin{align*}
    \frac{E_{n,d}'}{\phi_d(n)} = O\left(n^{-\frac2{d-1}} \right).
\end{align*}
Collecting these estimates proves the theorem. \qed

When $d$ increases beyond the range \eqref{d-rg1}, the error term
$f_d(n)$ (see \eqref{End-Rnd}) is no more negligible, and a more
delicate analysis is needed.
\begin{thm}[Uniform asymptotic estimate in the critical range]
\label{thm:ud2}
If
\begin{align} \label{d-rg2}
    \frac d{(\log n)^{1/3}}\to\infty\quad
    \text{and}\quad d\le 2\sqrt{\frac{\log n}
    {W\left(\frac{4\log n}{(e\log 2)^2}\right)}},
\end{align}
then, with $\rho := \frac{d}{e n^{1/d^2}}$,
\begin{align} \label{dm1-dom2}
    \mathbb{E}[M_{d,d-1}(n)] =
    \frac{n^{-\frac1{d-1}}}{d-1}\,\Gamma\left(\frac1{d-1}\right)^d
    \left(\frac1{2-e^{-\rho}}+
    O\left(\frac{\rho (\rho+1)e^{-\rho}}{(2-e^{-\rho})^3}
    \left(\frac1d+\frac{\log n}{d^3}\right)\right)\right),
\end{align}
uniformly in $d$ for large $n$.
\end{thm}
The proof of this theorem is very long and is thus relegated in
Appendix B. The crucial step is to prove an asymptotic estimate
for $f_d(n)$ by an inductive argument by deriving first a recurrence
of the form
\[
    f_d(n) = g_d(n) + \Phi[f_d](n) + \text{smaller order
    terms},
\]
where
\begin{align*} 
    g_d(n) := \sum_{1\le j\le d-2}\binom{d}{j}
    (-1)^{j-1} (d-1-j)^{j-1} \Gamma\left(\tfrac1{d-1-j}\right)^{d-j}
    n^{\frac1{d-1}-\frac{1}{d-1-j}} ,
\end{align*}
and $\Phi$ is an operator defined by
\[
    \Phi[f_d](n) := \sum_{1\le j\le d-2}\binom{d}{j}
    (-1)^j n^{\frac1{d-1}-\frac{1}{d-1-j}}
    \int_{(1,\infty)^j}\left(v_1\cdots v_j
    \right)^{-1-\frac1{d-1-j}} f_{d-j}(nv_1\cdots v_j)\text{d}\v{v}.
\]
Then \eqref{dm1-dom2} follows from iterating the operator and a
careful analysis of the resulting sums.

\begin{cor} If $d$ is of the form
\[
    d = \left\lfloor\sqrt{\frac{2\log n}{W(2\log n)-2v-2}}
    \right \rfloor,
\]
then
\begin{align} \label{cor-ph-tr}
    \frac{\mathbb{E}[M_{d,d-1}(n)]}
    {\frac{n^{-\frac1{d-1}}}{d-1}\,
    \Gamma\left(\frac1{d-1}\right)^d}
    \sim \left\{\begin{array}{ll}
        1, &\text{if }v\to-\infty;\\
        \frac1{2-e^{-e^v}},&\text{if }v=O(1);\\
        \frac12, &\text{if } v\to\infty.
    \end{array} \right.
\end{align}
\end{cor}
\pf  Observe that
\[
    \rho = \frac{d}{e n^{1/d^2}}
    = e^v\left(1+O\left(\frac{1+|v|}{W(2\log n)}\right)
    \right).
\]
Thus \eqref{cor-ph-tr} follows from this and \eqref{dm1-dom2}. 
\qed

Combining the ranges \eqref{d-rg1} and \eqref{d-rg2} of the two
estimates \eqref{dm1-dom} and \eqref{dm1-dom2}, we see that
\begin{cor} If
\[
    3\le d\le 2\sqrt{\frac{\log n}{W(4e^{-2}\log n)}},
\]
then
\[
    \mathbb{E}[M_{d,d-1}(n)] \sim
    \frac{1}{2-e^{-\rho}}\cdot
    \frac{n^{-\frac1{d-1}}}
    {d-1}\Gamma\left(\frac1{d-1}\right)^d,
\]
uniformly in $d$.
\end{cor}
We conclude from these estimates that $\mathbb{E}[M_{d,d-1}(n)]$ is,
modulo a constant term, very well approximated by
$\frac{n^{-\frac1{d-1}}}{d-1}\Gamma\left(\frac1{d-1}\right)^d$.

\medskip

\noindent \textbf{Remark.} A similar analysis as that for
\eqref{dm1-dom} leads to ($L_{d,k}(n,j)$ is defined in
Section~\ref{sec:layers})
\begin{align} \label{Ldd}
    \mathbb{E}[L_{d,d-1}(n,j)]
    \sim c_{d,j} n^{-\frac1{d-1}},
\end{align}
for each finite integer $j\ge0$, where
\begin{align*}
    c_{d,j}
    &:= \frac1{(d-1)j!}
    \int_{\mathbb{R}_+^d}(v_1+\cdots+v_d)^j
    e^{-(v_1+\cdots +v_d)}(v_1\cdots v_d)^{-\frac{d-2}{d-1}}
    \text{d}\v{v}\\
    &= \frac1{d-1}\,\Gamma\left(\frac1{d-1}\right)^d
    \binom{j+\frac1{d-1}}{j},
\end{align*}
uniformly when $\frac{2\log n}{d^2}-W(2\log n)\to\infty$ and
$j=o\left(n^{\frac{1-\varepsilon}{d}}\right)$, $\varepsilon\in(0,1)$.
The consideration for larger $d$ as for \eqref{dm1-dom2} is similar.

\section{Threshold phenomenon for $\mb{E}[M_{d,d-1}(n)]$ when
$d\to\infty$} \label{sec:threshold}

With the asymptotic estimates \eqref{dm1-dom} and \eqref{dm1-dom2}
we derived in the previous section, we prove in this section a less
expected threshold phenomenon for the expected number of
$(d-1)$-dominant skylines $\mb{E}[M_{d,d-1}(n)]$ (in random samples
from $d$-dimensional hypercube) when $d-1$ is near
$\sqrt{\frac{2\log n}{W(2\log n)}}$.

\begin{thm}[Threshold phenomenon] Let
\begin{align} \label{d-large}
    d_0 =d_0(n):= \tr{\sqrt{\frac{2\log n}{W(2\log n)}}}
    +1,
\end{align}
where $W$ denotes the Lambert-W function. Then the expected number
of ($d-1$)-dominant skyline points satisfies
\begin{align} \label{EMkn-large}
    \lim_{n\to\infty}\mb{E}[M_{d,d-1}(n)] \to
    \left\{\begin{array}{ll}
        0, &\text{if } d< d_0;\\
        \infty,& \text{if } d>d_0+1.
    \end{array} \right.
\end{align}
If $d=d_0$, then $\lim_{n\to\infty} \mb{E}[M_{d,d-1}(n)]$ does not
exist and is oscillating between $0$ and
$\frac{e^{-\gamma}}{2-e^{-e^{-1}}}$
\begin{align} \label{n-ii2}
    \mb{E}[M_{d,d-1}(n)] \sim
    \frac{e^{-\gamma}}{2-e^{-e^{-1}}}\,
    \varphi_0\left(\sqrt{\frac{2\log n}
    {W(2\log n)}}\right),
\end{align}
where $\varphi_0(x)$ is a bounded oscillating function of $x$
defined by
\[
    \varphi_0(x) := e^{-\{x\}} x^{-2\{x\}}.
\]
If $d=d_0+1$, then $\lim_{n\to\infty} \mb{E}[M_{d,d-1}(n)]$ does not
exist and is oscillating between $\frac{e^{-\gamma}}{2-e^{-e^{-1}}}$
and $O\left(\frac{\log n}{\log\log n}\right)$
\begin{align} \label{n-ii3}
    \mb{E}[M_{d,d-1}(n)] \sim
    \frac{e^{-\gamma}}{2-e^{-e^{-1}}}\,
    \varphi_1\left(\sqrt{\frac{2\log n}
    {W(2\log n)}}\right),
\end{align}
where $\varphi_1(x)$ is an oscillating function of $x$ defined by
\[
    \varphi_1(x) := e^{1-\{x\}} x^{2-2\{x\}}.
\]
\end{thm}
\pf  
By monotonicity, it suffices to examine the asymptotic behavior
of $\mathbb{E}[M_{d,d-1}(n)]$ for $d$ near $d_0$. Observe that if
\[
    d = d_0+m = \sqrt{\frac{2\log n}{W_n}}-\tau_n
    +m + 1,
\]
where $m$ is an integer and $\tau$ denotes the fractional part of
$\sqrt{\frac{2\log n}{W(2\log n)}}$, namely,
\[
    \tau_n :=
    \left\{\sqrt{\frac{2\log n}{W_n}}\right\}
    = \sqrt{\frac{2\log n}{W_n}} -
    \tr{\sqrt{\frac{2\log n}{W_n}}},
\]
then
\[
    \rho = \frac{d}{e n^{1/d^2}} = e^{-1}\left(1+O
    \left(\frac{W_n^{\frac32}|m+\tau_n|}{\sqrt{\log n}}
    \right)\right)\to e^{-1},
\]
where, here and throughout the proof, $W_n := W(2\log n)$. Thus for
bounded $m$
\[
    \frac1{2-e^{-\rho}}\to\frac1{2-e^{-e^{-1}}}.
\]

On the other hand, by \eqref{dm1-dom2} and the asymptotic estimate
$\Gamma(x)= x^{-1}-\gamma+O(x)$ as $x\to0$, where $\gamma$ denotes
the Euler constant, we see that
\begin{align*}
    \frac{n^{-\frac1{d-1}}}{d-1}\Gamma\left(\frac1{d-1}\right)^d
    &= e^{-\gamma+m-\tau_n}\left(\frac{2\log n}{W_n}
    \right)^{m-\tau_n}\left(1+O\left(
    \frac{W_n^{\frac32}(m+\tau_n+1)^2}{\sqrt{\log n}}
    \right)\right)\\
    &\left\{\begin{array}{ll}
        \to 0, &\text{if } m\le -1;\\
        \sim e^{-\gamma}\varphi_0\left(\sqrt{\frac{2\log n}
        {W_n}}\right),
        &\text{if } m=0 ;\\
        \sim e^{-\gamma}\varphi_1\left(\sqrt{\frac{2\log n}
        {W_n}}\right),
        &\text{if } m=1 ;\\
        \to \infty, &\text{if } m\ge 2.
    \end{array} \right.
\end{align*}
This proves \eqref{EMkn-large}, \eqref{n-ii2} and \eqref{n-ii3}. It
remains to consider more precisely the behavior of $\varphi_0(x)$
and $\varphi_1(x)$.

Obviously, by definition, $\varphi_0(x)\in (0,1]$ and
$\varphi_1(x)\in [1,\infty)$ because $\{x\}\in[0,1)$ for
$x\in\mathbb{R}_+$. If $\{x\}=0$, then $\varphi_0(x)=1$; more
generally,
\[
    \varphi_0(x) \to \left\{\begin{array}{ll}
        1, & \text{if }\{x\}\log x = o(1);\\
        0, & \text{if }\{x\}\log x \to\infty.
    \end{array}\right.
\]
On the other hand,
\[
    \varphi_1(x) \to \left\{\begin{array}{ll}
        1, & \text{if }(1-\{x\})\log x = o(1);\\
        \infty, & \text{if }(1-\{x\})\log x \to\infty.
    \end{array}\right.
\]

We now prove that
\begin{align} \label{phi-zero}
    \tau_n=0  \text{ if and only if } n= i^{i^2} \; (i\ge2).
\end{align}
First, if $n=i^{i^2}$, then $2\log n=2i^2\log i$ and the positive
solution to the equation (see \eqref{lambert-w})
\[
    W_ne^{W_n} = 2i^2\log i,
\]
is given by $W_n= 2\log i$, as can be easily checked. Thus
\begin{align} \label{ii2}
    \sqrt{\frac{2\log n}{W_n}}=i \qquad(i\ge2).
\end{align}
Conversely, if the relation \eqref{ii2} holds, then the positive
solution to the equations
\[
    \frac{2\log n}{W_n} = i^2,
    \text{ and } W_ne^{W_n} = 2\log n,
\]
is given by $n=i^{i^2}$. This proves \eqref{phi-zero}.

It follows particularly, by \eqref{dm1-dom2}, that
\[
    \lim_{i\to\infty} \mathbb{E}[M_{i,i-1}]\left(i^{i^2}\right)
    = \frac{e^{-\gamma}}{2-e^{-e^{-1}}}.
\]
This completes the proof of the theorem. \qed

The function $d_0$ of $n$ on the right-hand side of \eqref{d-large}
grows extremely slowly. Let $a_i := i^{i^2}$ with $a_1:=2$. Then
$d=i+1$ for $a_i\le n<a_{i+1}$, which is small for almost all
practical sizes of $n$
\[
    d_0 = \left\{\begin{array}{ll}
        2, & \text{if } 2\le n\le 15;\\
        3, & \text{if } 16\le n\le 19682;\\
        4, & \text{if } 19683 \le n\le 42949\,67295;\\
        5, & \text{if } 42949\,67296\le n\le
        2.98\dots\times 10^{17};\\
        6, & \text{if } 2.98\dots\times 10^{17}
        \le n\le 1.03\dots\times 10^{28}.
    \end{array} \right.
\]
This partly explains why the asymptotic vanishing property of
$\mb{E}[M_{d,k}(n)]$ for large $n$ and fixed $d$ is ``invisible''
for moderate values of $n$.

Note that we did not replace the Lambert-W function in
\eqref{d-large} by its asymptotic expansion \eqref{Wx} so as to make
the expression more transparent, the reason being that no matter how
many terms of the asymptotic expansion of $W$ we use, the resulting
expression is never $o(1)$. This is because all terms in the
expansion are of orders in powers of $\log \log n$ and $\log\log
\log n$, and they are all much smaller than $\log n$ in the
numerator of the first term on the right-hand side of
\eqref{d-large}.

Extending the same analysis to other values of $k$ becomes more
difficult and messy except for $k=1$ for which we have
\[
    \mathbb{E}[M_{d,1}(n)]
    = n \int_{[0,1]^d} (x_1\cdots x_d)^{n-1} \text{d}\v{x}
    = n^{1-d}.
\]
Note that this always tends to zero no matter how large the value of
$d$ is.

On the other hand, for $1\le k\le d-1$, we can derive the more
precise estimate
\begin{align*}
    \mathbb{E}[M_{d,k}(n)] &=O\left(n\int_{[0,1]^d} \exp\left(-n
    \sum_{1\le j_1<\cdots<j_k\le d}x_{j_1}\cdots x_{j_k}\right)
    \text{d} \v{x}\right)\\
    &=O\left( n^{1-\frac dk}\right).
\end{align*}
However, a more precise uniform asymptotic approximation (in $n, d$,
and $k$) is less obvious and describing the corresponding threshold
phenomena if any for other values of $k$ also remains unclear.
Intuitively, the asymptotic vanishing property is expected to hold
as long as $k \ge d/2$ no matter $d$ is finite or growing with $n$
because the probability of a $k$-dominance for a random pair of
points is larger than one half, meaning that it is less likely to
find $k$-dominant skyline in such a case.

\section{Expected number of dominant cycles}
\label{sec:nc}

The asymptotic zero-infinity property can be viewed from another
different angle by examining the \emph{number of dominant cycles}.

\medskip

\noindent \textbf{Definition.} We say that $m$ points
$\{\v{p}_1,\dots,\v{p}_m\}$ form a $k$-dominant cycle (of length
$m$) if $\v{p}_i$ $k$-dominates $\v{p}_{i+1}$ for $i=1, \dots,m-1$
and $\v{p}_m$ $k$-dominates $\v{p}_1$.

\medskip

Roughly, the number of $k$-dominant cycles is inversely proportional
to the number of $k$-dominant skylines. Note that by transitivity
there is no cycle when $k=d$. Thus the number of cycles seems a
better measure to clarify the structure of $k$-dominant skylines.
However, the general configuration of the cycle structure is very
complicated. We contend ourselves in this section with the
consideration of cycles of length $d$ when $k=d-1$.

\begin{lmm} Let $C_{n,d}$ denote the number of $(d-1)$-dominant
cycles of length $d$ in a random sample of $n$ points uniformly and
independently chosen from $[0,1]^d$. Then the expected value of
$C_{n,d}$ satisfies
\begin{align} \label{ECnd}
    \mathbb{E}[C_{n,d}] = \binom{n}{d} \frac{d!^{2-d}}{d}.
\end{align}
\end{lmm}
\pf  
Since the total number of cycles of length $d$ is given by
$\binom{n}d\frac{d!}{d}$, we see that
\[
    \mathbb{E}[C_{n,d}] = \binom{n}d\frac{d!}{d}
    \mathbb{P}\left(\{\v{p}_1,\dots,\v{p}_d\} \text{ form
    a $(d-1)$-dominant cycle of length $d$} \right).
\]
Assume that $\{\v{p}_1,\dots,\v{p}_d\}$ form a $(d-1)$-dominant
cycle of length $d$. Let
\[
    \v{p}_i = (p_{i,1},\dots,p_{i,d}) \qquad(i=1,\dots,d).
\]
Then for each coordinate $j$, there exists an $\ell$ such that
\[
    p_{1,j}>p_{2,j}>\cdots>p_{\ell, j}, \quad
    p_{\ell, j}<p_{\ell+1,j}, \quad
    p_{\ell+1,j}>\cdots>p_{d,j}>p_{1,j},
\]
and the $\ell$'s are all distinct ($d!$ cases). Thus the probability
of the event that $\{\v{p}_1,\dots,\v{p}_d\}$ form a
$(d-1)$-dominant cycle is given by
\[
    \frac{d!}{d!^d},
\]
from which (\ref{ECnd}) follows. \qed

In particular, we see that
\[
    \mathbb{E}[C_{n,2}] = \frac{n(n-1)}{4},
\]
which means that half of the pairs are cycles, rendering the
$1$-dominant skylines less likely to occur. The first few other
$\mathbb{E}[C_{n,d}]$ are given by
\begin{align*}
    \{\mathbb{E}[C_{n,d}] \}_{d\ge3}
    &= \left\{\tfrac{n(n-1)(n-2)}{108},
    \tfrac{n(n-1)(n-2)(n-3)}{55296},
    \tfrac{n(n-1)(n-2)(n-3)(n-4)}{1036800000},\right.\\
    &\qquad \left.
    \tfrac{n(n-1)(n-2)(n-3)(n-4)(n-5)}{1160950579200000},\dots
    \right\}.
\end{align*}
We see that the denominator grows very fast and we expect another
type of threshold phenomenon.

Let
\[
    d_1 := \tr{\frac{\log n}{W(e^{-1}\log n)}+\tfrac12},
\]
and $\tau_n$ denote the fractional part of $\frac{\log
n}{W(e^{-1}\log n)}+\tfrac12$. Also let
\begin{align*}
\begin{split}
    \upsilon(t) &:= \frac{1+\frac12\log 2\pi}{W+1}
    + \frac W{(\log n)(W+1)}\Biggl(t
    \\&\qquad \left. -
    \frac{12W^3+(35-12\log2\pi)W^2+(34-24\log2\pi)W+
    23+(\log2\pi)^2}{24(W+1)^3}\right),
\end{split}
\end{align*}
where $t\in\mathbb{R}$ and $W$ represents $W(e^{-1}\log n)$. Note
that $W$ is of order $\log\log n$.
\begin{thm} The expected number of $(d-1)$-dominant cycles of
length $d$ satisfies
\begin{align*}
    \lim_{n\to\infty}\mathbb{E}[C_{n,d}]
    \to \left\{\begin{array}{ll}
        \infty, & \text{if }2\le d<d_1;\\
        0, & \text{if } d>d_1.
    \end{array}
    \right.
\end{align*}
When $d=d_1$, we can write $\tau_n=\upsilon(t)$; then
\begin{align} \label{dd1}
    \lim_{n\to\infty}\mathbb{E}[C_{n,d}]
    \left\{\begin{array}{ll}
        \to 0& \text{if }t\to-\infty;\\
        \sim e^t, & \text{if }t=O(1);\\
        \to\infty,&\text{if }t\to\infty.
    \end{array} \right.
\end{align}
\end{thm}
\pf  Write
\[
    d = d_1 - m = \frac{\log n}{W(e^{-1}\log n)} + \tfrac12
    - v,
\]
where $v=m+\tau_n$. Then a straightforward calculation using
\eqref{ECnd} and Stirling's formula gives
\begin{align*}
    \frac1d\log \mathbb{E}[C_{n,d}]
    &= v\left(W(e^{-1}\log n)
    +1\right)
    - 1-\tfrac12\,\log 2\pi \\
    &\qquad\qquad +
    O\left(\frac{W(e^{-1}\log n)^2+(v^2+1)W(e^{-1}\log n)}
    {\log n}\right).
\end{align*}
Thus $\mathbb{E}[C_{n,d}]\to\infty$ if $m\ge1$ and
$\mathbb{E}[C_{n,d}]\to-\infty$ if $m\le -1$. When $m=0$
($v=\tau_n$), this asymptotic expansion is insufficient and we need
more terms.  If $v=\tau_n=\upsilon(t)$, then the same calculation as
above gives
\[
    \mathbb{E}[C_{n,d}] = e^t\left(1+O\left(
    \frac{W^2+1}{\log n}\right)\right).
\]
This implies \eqref{dd1}. \qed

Let
\[
    a_i := \left\lfloor\left(\tfrac{i-\frac12}{e}\right)^{i-\frac12}
    \right\rfloor+1\qquad(i\ge1).
\]
Then
\[
    d_1=d_1(n) = i \text{ if } a_i\le n<a_{i+1}.
\]
The first few values of $a_i$ are given as follows. 
\[
    \begin{tabular}{|c||c|c|c|c|c|c|c|c|c|} \hline
    $i$ & $4$ & $5$ & $6$ & $7$ & $8$ & $9$ &$10$
    & $11$ & $12$\\ \hline
    $a_i$ & $3$ & $10$ & $49$ & $290$ & $2022$ & $16165$
    &$145405$ & $1453435$ & $15982276$\\ \hline
    \end{tabular}
\]

\section{A uniform lower bound for $\mathbb{E}[M_{d,k}(n)]$}
\label{sec:ub}

The convergence rate in \eqref{a1} is very slow if $d$ is large and
$ k $ is close to $d$. It is interesting to characterize the
transition of $M_{d,k}(n)$ from zero to $ n$ as $k$ increases under
the condition that $d$ and $n$ are fixed. However, the exact
characterization is not easy, so we derive instead a lower bound
that provides a good approximation to the real transition.

\begin{thm}[Uniform lower bound in $d, k$ and $n$] Define
\[
    \beta_{d,k}:=\sum_{0\le j\le d-k}\binom{d}{j}2^{-d}.
\]
Then, for $n\ge1$ and $1\le k\le d-1$,
\begin{equation}
    \mathbb{E}[M_{d,k}(n)]
    \ge n  I_n(\beta_{d,k}), \label{t3}
\end{equation}
where
\[
    I_n(x) := x\int_{x}^{1}t^{-2}
    \left(1-t\right)^{n-1}\mathrm{d}t.
\]
\end{thm}
\pf  
Select two random points $\v{x},\v{y}$ uniformly and
independently in $ [0,1]^{d}$. Obviously,
\[
    \mathbb{P}\left( \v{x}\text{ $k$-dominates }
    \v{y}\right) =\beta_{d,k}.
\]
On the other hand, by definition, $\mathbb{P} \left( \v{x}\text{
$k$-dominates }\v{y} \right) = \int_{[0,1]^{d}}\left| B_{k}(\v{x})
\right| \mathrm{d}\v{x}$. Thus
\[
    \int_{\lbrack 0,1]^{d}}\left| B_{k}(\v{x})\right|
    \mathrm{d} \v{x}=\beta_{d,k}.
\]
Let
\[
    F(t)=\left| \left\{ \v{x}\in [0,1]^{d}
    :\left| B_{k}(\v{x})\right|
    \le t\right\} \right| ,
\]
be the distribution function of $|B_{k}(\v{x})|$. By Markov
inequality
\[
    t\left( 1-F(t)\right)
    \le \int_{\lbrack 0,1]^{d}}\left|
    B_{k}(\v{x})\right|\mathrm{d} \v{x}
    \qquad(t\in(0,1)) .
\]
Thus
\[
    F(t)
    \ge 1-\frac{\int_{[0,1]^{d}}\left|
    B_{k}(\v{x})\right|\mathrm{d}\v{x}}{t}
    =1-\frac{\beta_{d,k}}{t}.
\]
Define
\[
    G(t):=\max\left\{1-\frac{\beta_{d,k}}{t}, 0
    \right\}.
\]
Then $F(t)\ge G(t)$. Now
\begin{equation} \label{tm1}
    \mathbb{E}[M_{d,k}(n)]
    =n\int_{[0,1]^{d}}\left( 1-\left|B_{k}(\v{x})\right|
    \right)^{n-1}\mathrm{d}\v{x}
    =n\int_{0}^{1}\left( 1-t\right)^{n-1}\frac{\mathrm{d}F(t)}
    {\mathrm{d}t} .
\end{equation}
Since the integral on the right-hand side of \eqref{tm1} becomes
smaller if the distribution function $F(t)$ is replaced by $G(t)$,
we have
\begin{align*}
    \mathbb{E}[M_{d,k}(n)]
    \ge n\int_0^1 (1-t)^{n-1}
    \frac{\mathrm{d}G(t)}{\mathrm{d}t} ,
\end{align*}
from which \eqref{t3} follows. \qed

A useful, convergent asymptotic expansion for $I_n(x)$, derived by
successive integration by parts, is as follows.
\begin{align*}
    I_n(x) &= \sum_{j\ge0}
    \frac{(-1)^j (j+1)!}{n(n+1)\cdots(n+j)}
    \,x^{-j-1}(1-x)^{n+j}\\
    &= \frac{(1-x)^n}{nx} - \frac{2(1-x)^{n+1}}{n(n+1)x^2}
    + \cdots,
\end{align*}
as long as $x\gg1/n$. In particular, $I_n(x)\to0$ in this range of
$x$. If $xn\to c>0$, then
\[
    I_n(x) \to c\int_c^\infty u^{-2} e^{-u} \text{d}u,
\]
the latter tending to $1$ as $c$ approaches zero.

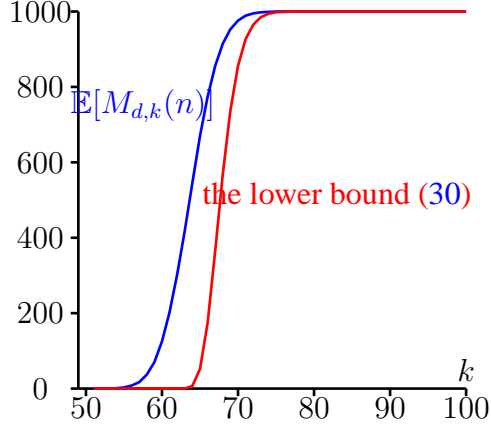
\begin{figure}[tbp]
\centering
\begin{tikzpicture}[scale=0.5]
\draw(10.20,.50) node{$k$};
\draw[line width=1pt](0,0)--(10.20,0); \foreach \x/\k in
{0.20/50,2.20/60,4.20/70,6.20/80,8.20/90,10.20/100} {%
\draw[line width=1pt](\x,-0.2)--(\x,0);
\draw(\x,-0.5)circle(0pt)node[black]{$\k$}; }%
\foreach \x in {} \draw[line width=1pt](\x,-0.1)--(\x,0);%
\draw[line width=1pt](0,0)--(0,10.00);
\foreach \y/\k in {0.00/0,2.00/200,4.00/400,
6.00/600,8.00/800,10.00/1000} { \draw[line
width=1pt](-0.2,\y)--(0,\y);
\draw(-1,\y)circle(0pt)node[black]{$\k$}; } \draw[line
width=1pt,color=black,mark=o,mark size=2pt,draw=blue] plot
coordinates{(0.40,0.00)(0.60,0.00)(0.80,0.00)(1.00,0.01)
(1.20,0.03)(1.40,0.08)(1.60,0.17)(1.80,0.37)(2.00,0.70)
(2.20,1.25)(2.40,2.03)(2.60,3.03)(2.80,4.19)(3.00,5.46)
(3.20,6.70)(3.40,7.74)(3.60,8.56)(3.80,9.15)(4.00,9.53)
(4.20,9.76)(4.40,9.89)(4.60,9.95)(4.80,9.98)(5.00,9.99)
(5.20,10.00)(5.40,10.00)(5.60,10.00)(5.80,10.00)(6.00,10.00)
(6.20,10.00)(6.40,10.00)(6.60,10.00)(6.80,10.00)(7.00,10.00)
(7.20,10.00)(7.40,10.00)(7.60,10.00)(7.80,10.00)(8.00,10.00)
(8.20,10.00)(8.40,10.00)(8.60,10.00)(8.80,10.00)(9.00,10.00)
(9.20,10.00)(9.40,10.00)(9.60,10.00)(9.80,10.00)(10.00,10.00)
(10.20,10.00)}node[right]{{\color{blue} }};
\draw(1.7,7.5)circle(0pt)node[blue]{$\mb{E}[M_{d,k}(n)]$};
\draw[line
width=1pt,color=black,mark=pentagon,mark size=2pt,draw=red] plot
coordinates{(0.40,0.00)(0.60,0.00)(0.80,0.00)(1.00,0.00)
(1.20,0.00)(1.40,0.00)(1.60,0.00)(1.80,0.00)(2.00,0.00)
(2.20,0.00)(2.40,0.00)(2.60,0.00)(2.80,0.00)(3.00,0.07)
(3.20,0.52)(3.40,1.74)(3.60,3.65)(3.80,5.69)(4.00,7.38)
(4.20,8.56)(4.40,9.27)(4.60,9.65)(4.80,9.85)(5.00,9.94)
(5.20,9.98)(5.40,9.99)(5.60,10.00)(5.80,10.00)(6.00,10.00)
(6.20,10.00)(6.40,10.00)(6.60,10.00)(6.80,10.00)(7.00,10.00)
(7.20,10.00)(7.40,10.00)(7.60,10.00)(7.80,10.00)(8.00,10.00)
(8.20,10.00)(8.40,10.00)(8.60,10.00)(8.80,10.00)(9.00,10.00)
(9.20,10.00)(9.40,10.00)(9.60,10.00)(9.80,10.00)(10.00,10.00)
(10.20,10.00)} node[right]{{\color{blue} }};%
\draw(6.8,5.1) circle(0pt)node[red]{the lower bound \eqref{t3}};
\end{tikzpicture}
\caption{\emph{Simulation result of $\mathbb{E}[M_{d,k}(n)]$ and the
lower bound \eqref{t3} for $n=1000,d=100$ and $k$ from $50$ to
$100$.}}\label{fg1}
\end{figure}

We see that the transition of $I_n(x)$ from zero to one occurs at
$x\asymp n^{-1}$ (meaning that $x$ is of order proportional to
$n^{-1}$). In terms of $d$ and $k$, this arises when $d\to\infty$
and $\beta_{d,k}\asymp n^{-1}$. Now, by known estimate for binomial
distribution (see \cite{Hwang97} and the references cited there)
\[
    \beta_{d,k} \asymp (2\alpha-1)^{-1}
    d^{-1/2} 2^{-d}\alpha^{-\alpha d}
    (1-\alpha)^{-(1-\alpha)d},
\]
when $k\ge d/2+K\sqrt{d}$, where $\alpha := k/d$ and $K>1$ is a
constant. We deduce from this that the transition of
$I_n(\beta_{d,k})$ from zero to one occurs at $c\log n$ for some
$c\in(0,1)$. The exact location of this $c$ matters less since $I_n$
is simply a lower bound; see Figure~\ref{fg1}.

\section{Conclusions}
\label{sec:fin}

While the notion of $k$-dominant skyline appeared as a natural means
of solving the abundance of skyline, its use in diverse contexts has
to be carefully considered, in view of the results we derived in
this paper. We summarize our findings and highlight suggestions for
possible practical uses.

The asymptotic results we derived in this paper are either of a
vanishing type or of a blow-up nature; briefly, they are either zero
or infinity when the sample size goes unbounded, making the
selection of representative points more subtle. The expected number
of $k$-dominant skyline points approaches zero under either of the
following situations.
\begin{itemize}
\item Hypercube: both $d$ and $k<d$ bounded;
\item Simplex: both $d$ and $k<d$ bounded;
\item Hypercube: extending the $k$-dominant skyline to
the dominance by a cluster of $j$ points with both $d$ and $k$
bounded.
\end{itemize}
In all cases, zero appears as the limit when $n\to\infty$. However,
for practical purposes, $n$ is always finite, and thus the above
limit results become less useful from a computational point of view.
One needs asymptotic estimates that are uniform in $d$, $k$ and $n$.
But such results are often very difficult. The uniform asymptotic
approximation \eqref{dm1-dom} we obtained leads to several
interesting consequences, including particularly the threshold
phenomenon \eqref{EMkn-large}.

We conclude this paper by showing how the asymptotic results we
derived above can be applied in more practical situations. Assume
that our sample is of size, say $n=10^4$ or $n=10^5$, and the
dimensionality $d$ is in the range $\{4,5,6,7,8\}$ (smaller $d$ may
result in more biased inferences while larger $d$ will yield too
many skyline points). We also assume that our data set is
sufficiently random and can be modeled by the hypercube model. If
our aim is to choose a reasonably small number of candidates for
further decision making, then how can our asymptotic estimates help?

First, for this range of $n$ and $d$, the expected numbers of
skyline points can be easily computed by the recurrence relation
(see \cite{BDHT05})
\[
    \mu_{n,d} = \frac1{d-1}\sum_{1\le j\le d-1}
    H_n^{(d-j)}\mu_{n,j} \qquad(d\ge2),
\]
where $\mu_{n,d}:= \mb{E}[M_{d,d}(n)]$, $H_n^{(a)} := \sum_{1\le
j\le n}j^{-a}$ are the harmonic numbers and $\mu_{n,1} := 1$, and
are given approximately by
\[
    \{164.7, 426.3, 902.7, 1633.1,2603\} \qquad(n=10^4;d=4,5,6,7,8),
\]
and
\[
    \{304.9, 955.8, 2432.1, 5239.4,9845\} \qquad(n=10^5;d=4,5,6,7,8),
\]
which are often too many for further consideration. So we turn to
$(d-1)$-dominant skyline and estimate their numbers by our
asymptotic approximations. However, both Theorems~\ref{thm:ud} and
\ref{thm:ud2} have poor error terms, and a better numerical
approximation to $\mb{E}[M_{d,d-1}(n)$ for most moderately values
of $n$ and $d$ is given by
\[
    \phi_d(n)-g_d(n)= \sum_{0\le j\le d-2}\binom{d}{j}
    (-1)^j (d-1-j)^{j-1} \Gamma\left(\tfrac1{d-1-j}\right)^{d-j}
    n^{\frac1{d-1}-\frac{1}{d-1-j}}.
\]
We thus obtain, for example, the following numerical values
\begin{align*}
    \mb{E}[M_{d,d-1}(10^4)]\approx
    \begin{tabular}{|c||c|c|c|c|c|} \hline
    $d$ & $4$ & $5$ & $6$ & $7$ & $8$ \\ \hline\hline
    $\phi_d(n)-g_d(n)$ & $0.61$ & $5.06$ & $24.85$ & $88.90$ & $243.96$
    \\ \hline
    Monte Carlo & $0.57$ & $4.82$ & $23.98$ & $83.89$ &
    $226.65$ \\ \hline
    \end{tabular}
\end{align*}
and
\begin{align*}
    \mb{E}[M_{d,d-1}(10^5)]\approx
    \begin{tabular}{|c||c|c|c|c|c|} \hline
    $d$ & 4 & 5 & 6 & 7 & 8\\ \hline\hline
    $\phi_d(n)-g_d(n)$ & 0.31& 3.69 & 24.94& 115.31 & 404.7\\ \hline
    Monte Carlo & 0.29 & 3.61 & 24.38 & 111.79 & 386.08\\ \hline
    \end{tabular}
\end{align*}
From these tables, one can choose a suitable $d$ according to the
need of practical uses. Here we also see the characteristic property
of the skylines, either very few or very many points.

Our Monte Carlo simulations are carried out by a three-phase
algorithm (extending our two-phase maxima-finding one in
\cite{CHT11}) for finding the $k$-dominant skylines. Briefly, the
first two phases are modified from the algorithms presented
in \cite{CHT11} and the last phase removes all cycles.

\section*{Acknowledgements}
We thank Yuliy Baryshnikov for pointing out the references
\cite{BO96} and \cite{Orlova91}.

\section*{Appendix A. Error analysis: $d\le \sqrt{\frac{2\log n}
{W(2\log n)+K}}$}

Recall that $N := n^{\frac1{d-1}}$ and consider the integral
\begin{align*}
    f_d(n) = \left(\int_{\mathbb{R}_+^d}-\int_{[0,N]^d}\right)
    e^{-y_1\cdots y_d \left(\frac1{y_1}+\cdots+\frac1{y_d}
    \right)} \text{d}\v{y}
    = \sum_{1\le j\le d}\binom{d}{j}(-1)^{j-1} \phi_{d,j}(n),
\end{align*}
where
\begin{align}\label{phi-djn}
    \phi_{d,j}(n) := \int_{[0,N]^{d-j}
    \times (N,\infty)^j} e^{-y_1\cdots y_d
    \left(\frac1{y_1}+\cdots+\frac1{y_d}
    \right)} \text{d}\v{y}.
\end{align}
So our $\phi_d(n)=\frac1{d-1}\Gamma\left(\frac1{d-1}\right)^d$
corresponds to $\phi_{d,0}(n)$; see \eqref{End-Rnd}.
\begin{prop} \label{prop-1}
Let $d\ge3$ satisfies $\frac{2\log n}{d^2}-W(2\log n)\to\infty$.
Then
\begin{align}\label{Err-Bd}
    f_d(n) = O\left(\phi_d(n)
    d N^{-\frac1{d-2}}\right),
\end{align}
uniformly in $d$.
\end{prop}
\pf  We first prove that uniformly for $1\le j\le d$,
\begin{align} \label{Endj}
    \phi_{d,j}(n) = O\left(\Gamma\left(\tfrac1{d-2}\right)^{d-1}
    N^{-\frac{j}{d-2}}\right).
\end{align}
Consider first the range $1\le j\le d-2$. By extending the
integration ranges and then carrying out the changes of variables
$y_\ell \mapsto Nv_{d-\ell+1}$ for $d-j+1\le \ell \le d$, we obtain
the bounds
\begin{align*}
    \phi_{d,j}(n)
    &= N^j\int_{(1,\infty)^j}\!\!\int_{[0,N]^{d-j}}
    e^{-N^j v_1\cdots v_j y_1\cdots y_{d-j}
    \left(\frac1{y_1}+\cdots+\frac1{y_{d-j}}+
    \frac1{Nv_1}+\cdots+\frac1{Nv_j}
    \right)} \text{d}\v{y}\text{d}\v{v} \\
    &\le N^j\int_{(1,\infty)^j}\!\!\int_{\mathbb{R}_+^{d-j}}
    e^{-N^j v_1\cdots v_jy_1\cdots y_{d-j}
    \left(\frac1{y_1}+\cdots+\frac1{y_{d-j}}
    \right)} \text{d}\v{y}\text{d}\v{v}.
\end{align*}
By the change of variables $y_j \mapsto \lambda^{-\frac1{d-1}} x_j$
for $1\le j\le d$, we have, for $\lambda>0$,
\[
    \int_{\mathbb{R}_+^{d}}
    e^{-\lambda y_1\cdots y_{d}
    \left(\frac1{y_1}+\cdots+\frac1{y_{d}}\right)}\text{d}\v{y}
    = \frac{\Gamma\left(\tfrac1{d-1}\right)^{d}}
    {d-1}\,\lambda^{-\frac{d}{d-1}}\qquad(d\ge2).
\]
It follows that
\begin{align*}
    \phi_{d,j}(n) &\le \frac{\Gamma\left(\tfrac1{d-1-j}\right)^{d-j}}
    {d-1-j} \, N^{-\frac{j}{d-1-j}} \int_{(1,\infty)^j}
    (v_1\cdots v_j)^{-1-\frac1{d-1-j}}\text{d}\v{v}\\
    &= (d-1-j)^{j-1} \Gamma\left(\tfrac1{d-1-j}\right)^{d-j}
    N^{-\frac{j}{d-1-j}}\\
    &= O\left(\Gamma\left(\tfrac1{d-2}\right)^{d-1}
    N^{-\frac{j}{d-2}}\right),
\end{align*}
uniformly for $1\le j\le d-2$. The remaining two cases $j=d-1, d$
are much smaller; we start with $\phi_{d,d}(n)$. By the same
analysis used above, we have
\begin{align*}
    \phi_{d,d}(n) &= \int_{(N,\infty)^d} e^{-
    x_1\cdots x_d \left(\frac1{x_1}+\cdots+\frac1{x_{d}}\right)}
    \text{d}\v{x}\\
    &\le \int_{(N,\infty)^d} e^{-
    x_1\cdots x_d \left(\frac1{x_1}+\cdots+\frac1{x_{d-1}}\right)}
    \text{d}\v{x}\\
    &\le \int_{(N,\infty)^{d-1}} \frac{e^{-N
    x_1\cdots x_{d-1} \left(\frac1{x_1}+\cdots+\frac1{x_{d-2}}
    \right)}}{x_1\cdots x_{d-1}
    \left(\frac1{x_1}+\cdots+\frac1{x_{d-2}}\right)}
    \text{d}\v{x}.
\end{align*}
By the inequality
\begin{align} \label{ta-ineq}
    \int_N^\infty t^{-\alpha} e^{-\lambda t} \text{d}t
    \le \lambda^{-1}N^{-\alpha} e^{-\lambda N}
    \qquad(\alpha\ge0,\lambda>0),
\end{align}
we obtain
\begin{align*}
    \phi_{d,d}(n) &\le N^{-2} \int_{(N,\infty)^{d-2}}
    \frac{e^{-N^2
    x_1\cdots x_{d-2} \left(\frac1{x_1}+\cdots+\frac1{x_{d-2}}
    \right)}}{x_1\cdots x_{d-2}
    \left(\frac1{x_1}+\cdots+\frac1{x_{d-2}}\right)}
    \text{d}\v{x}\\
    &\le \cdots\\
    &\le N^{-2-4-\cdots-2(d-3)}\int_{(N,\infty)^2}
    \frac{e^{-N^{d-2}(x_1+x_2)}}{(x_1+x_2)^{d-2}}
    \text{d}\v{x}\\
    &= N^{-(d-2)(d-3)} \int_{2N}^\infty \frac{e^{-N^{d-2}w}}
    {w^{d-2}}(w-2N) \text{d}w\\
    &\le 2^{3-d} N^{-d^2+3d-1} e^{-2N^{d-1}}.
\end{align*}
Thus
\begin{align}\label{phi-ddn}
    \phi_{d,d}(n) = O\left(2^{-d} n^{-d+2+\frac1{d-1}} e^{-2n}\right).
\end{align}

Finally,
\begin{align*}
    \phi_{d,d-1}(n) &\le \int_{(N,\infty)^{d-1}} \frac{e^{-
    x_1\cdots x_{d-1}}}
    {x_1\cdots x_{d-1} \left(\frac1{x_1}+\cdots+\frac1{x_{d-1}}\right)}
    \,\text{d}\v{x}\\
    &\le \frac{1}{d-1}\int_{(N,\infty)^{d-1}}
    \frac{e^{-x_1\cdots x_{d-1} }}
    {(x_1\cdots x_{d-1})^{1+\frac1{d-1}}}
    \,\text{d}\v{x},
\end{align*}
by the inequality of arithmetic and geometric means
\[
    \frac1{d-1}\left(\frac1{x_1}+\cdots+\frac1{x_{d-1}}\right)
    \ge (x_1\cdots x_{d-1})^{\frac1{d-1}}.
\]
Applying successively the inequality \eqref{ta-ineq}, we obtain
\begin{align*}
    \phi_{d,d-1}(n) &\le \frac{N^{-1-\frac1{d-1}}}{d-1}
    \int_{(N,\infty)^{d-2}}
    \frac{e^{-Nx_1\cdots x_{d-1}}}
    {(x_1\cdots x_{d-2})^{2+\frac1{d-1}}}
    \,\text{d}\v{x}\\
    &\le \cdots\\
    &\le \frac{N^{-(d^2-2d+2)}}{d-1}
    e^{-N^{d-1}}.
\end{align*}
It follows that
\begin{align}\label{phi-dd1n}
    \phi_{d,d-1}(n) = O\left(d^{-1}
    n^{-d+1-\frac1{d-1}} e^{-n}\right).
\end{align}
We see that both $\phi_{d,d}(n)$ and $\phi_{d,d}(n)$ are much
smaller than the right-hand side of \eqref{Endj}.

The remaining case is when $d=2$. Obviously,
\[
    \phi_{2,1}(n) < \int_0^\infty \!\!\!\int_N^\infty e^{-y_1-y_2}
    \text{d}y_2\text{d}y_1 = e^{-N}.
\]

The upper bound \eqref{Err-Bd} then follows from summing
$\phi_{d,j}(n)$ for $j$ from $1$ to $d$ using \eqref{Endj}
\begin{align*}
    \sum_{1\le j\le d}\binom{d}j (-1)^{j-1} \phi_{d,j}(n)
    &= O\left(\Gamma\left(\tfrac1{d-2}\right)^{d-1}
    \sum_{j\ge1} \frac{d^j}{j!} N^{-\frac{j}{d-2}}\right)\\
    &= O\left(\Gamma\left(\tfrac1{d-2}\right)^{d-1}
    d N^{-\frac1{d-2}}\right),
\end{align*}
since $dN^{-\frac1{d-2}}\to0$ for $d$ in the range \eqref{d-rg1}.

It remains to estimate $R_d(n)$, which can be proved to be bounded
above by
\begin{align*}
    R_d(n) &= O\left(\frac{d}{N}\int_{\mathbb{R}_+^d}
    y_1\cdots y_d e^{-y_1\cdots y_d\left(\frac1{y_1}+\cdots
    +\frac1{y_d}\right)} \text{d} \v{y}\right)\\
    &= O\left(\frac{1}{N}\Gamma\left(\frac{2}{d-1}\right)^d \right);
\end{align*}
this proves \eqref{Rdn-ratio}. \qed

\section*{Appendix B. Proof of Theorem~\ref{thm:ud2}}

We prove Theorem~\ref{thm:ud2} in this Appendix. Our method of proof
consists in a finer evaluation of the integrals $\phi_{d,j}(n)$,
leading to a more precise asymptotic approximation to $f_d(n)$.

\begin{prop} Uniformly for $d$ in the range \eqref{d-rg2}
\begin{align}\label{fdn-asymp}
    f_d(n) \sim \frac{1-e^{-\rho}}{2-e^{-\rho}}\cdot
    \frac1{d-1}\Gamma\left(\frac1{d-1}\right)^d,
\end{align}
where $\rho := \frac{d}{en^{1/d^2}}$.
\end{prop}
\pf 
Consider again \eqref{phi-djn} and start with the changes of
variables $y_\ell \mapsto Nv_{d-\ell+1}$ for $d-j+1\le \ell \le d$,
\begin{align*}
    \phi_{d,j}(n)
    = N^j\int_{(1,\infty)^j}\!\!\int_{[0,N]^{d-j}}
    e^{-\lambda_{N,j}(\v{v})y_1\cdots y_{d-j}
    \left(\frac1{y_1}+\cdots+\frac1{y_{d-j}}+
    \frac1{Nv_1}+\cdots+\frac1{Nv_j}
    \right)} \text{d}\v{y}\text{d}\v{v},
\end{align*}
where $\lambda_{N,j}(\v{v}) := N^j v_1\cdots v_j$. Then we carry out
the change of variables
\[
    y_\ell \mapsto \lambda_{N,j}(\v{v})^{-\frac1{d-1-j}} x_\ell
    \qquad(1\le \ell \le d-j),
\]
and obtain
\[
    \phi_{d,j}(n) = \psi_{d,j}(n)+\omega_{d,j}(n),
\]
where
\begin{align*}
    \psi_{d,j}(n) &=
    N^{-\frac j{d-1-j}}\int_{(1,\infty)^j}\left(v_1\cdots v_j
    \right)^{-1-\frac1{d-1-j}}
    \int_{[0,N_0]^{d-j}}
    e^{-x_1\cdots x_{d-j}
    \left(\frac1{x_1}+\cdots+\frac1{x_{d-j}}
    \right)} \text{d}\v{x}\text{d}\v{v},
\end{align*}
with
\[
    N_0 := N^{\frac{d-1}{d-1-j}}(v_1\cdots v_j)^{\frac1{d-1-j}}
    = (n v_1\cdots v_j)^{\frac1{d-1-j}},
\]
and the error introduced is bounded above by
\begin{align*}
    \omega_{d,j}(n) &:= N^{-\frac j{d-1-j}}
    \int_{(1,\infty)^j}\left(v_1\cdots v_j
    \right)^{-1-\frac1{d-1-j}} \\
    &\qquad \times\int_{[0,N_0]^{d-j}}
    e^{-x_1\cdots x_{d-j}
    \left(\frac1{x_1}+\cdots+\frac1{x_{d-j}}
    \right)} \left(e^{-\frac{x_1\cdots x_{d-j}}{N_0}
    \left(\frac1{v_1}+\cdots+\frac1{v_j}\right)}-1\right)
    \text{d}\v{x}\text{d}\v{v} \\
    &= O\left(N^{-1-\frac {2j}{d-1-j}}
    \int_{(1,\infty)^j}\left(v_1\cdots v_j
    \right)^{-1-\frac2{d-1-j}} \left(\tfrac1{v_1}+\cdots
    +\tfrac1{v_j} \right)\right.\\
    &\qquad\qquad\qquad\left. \times\int_{\mathbb{R}_+^{d-j}}
    e^{-x_1\cdots x_{d-j}
    \left(\frac1{x_1}+\cdots+\frac1{x_{d-j}}
    \right)}x_1\cdots x_{d-j} \text{d}\v{x}\text{d}\v{v}
    \right)\\
    &= O\left(j 2^{-j}
    (d-1-j)^{j-2}\Gamma\left(\tfrac{2}{d-1-j}\right)^{d-j}
    N^{-1-\frac{2j}{d-1-j}}\right).
\end{align*}
Thus the total contribution of $\omega_{d,j}(n)$ to $f_d(n)$ is
bounded above by
\begin{align} \label{hdn}
\begin{split}
    h_d(n) &:=
    \sum_{1\le j\le d-2} \binom{d}{j}(-1)^{j-1}
    \omega_{d,j}(n) \\
    &\le \sum_{1\le j\le d-2}\binom{d}{j} j 2^{-j}
    (d-1-j)^{j-2}\Gamma\left(\tfrac{2}{d-1-j}\right)^{d-j}
    n^{\frac1{d-1}-\frac{2}{d-1-j}},
\end{split}
\end{align}
which will be seen to be of a smaller order.

\paragraph{The recurrence relation} Now
\begin{align*}
    \psi_{d,j}(n) &=
    N^{-\frac j{d-1-j}}\int_{(1,\infty)^j}\left(v_1\cdots v_j
    \right)^{-1-\frac1{d-1-j}} \int_{\mathbb{R}_+^{d-j}}
    e^{-x_1\cdots x_{d-j}
    \left(\frac1{x_1}+\cdots+\frac1{x_{d-j}}
    \right)} \text{d}\v{x}\text{d}\v{v} \\
    &\qquad - N^{-\frac j{d-1-j}}\int_{(1,\infty)^j}\left(v_1\cdots v_j
    \right)^{-1-\frac1{d-1-j}} f_{d-j}(nv_1\cdots v_j)\text{d}\v{v}\\
    &= (d-1-j)^{j-1} \Gamma\left(\tfrac1{d-1-j}\right)^{d-j}
    N^{-\frac{j}{d-1-j}} \\
    &\qquad - N^{-\frac j{d-1-j}}\int_{(1,\infty)^j}\left(v_1\cdots v_j
    \right)^{-1-\frac1{d-1-j}} f_{d-j}(nv_1\cdots v_j)\text{d}\v{v}.
\end{align*}
So we get the following recurrence relation.
\begin{lmm} The integrals $f_d(n)$ satisfy
\begin{align}
    f_d(n) &= g_d(n) + h_d(n) + \eta_d(n)\nonumber \\
    & \quad + \sum_{1\le j\le d-2}\binom{d}{j}
    (-1)^j n^{\frac1{d-1}-\frac{1}{d-1-j}}
    \int_{(1,\infty)^j}\left(v_1\cdots v_j
    \right)^{-1-\frac1{d-1-j}}
    f_{d-j}(nv_1\cdots v_j)\mathrm{d}\v{v},
    \label{fdn-rr}
\end{align}
for $d\ge3$, with the initial condition
\[
    f_2(n) = 2e^{-n}-e^{-2n},
\]
where $h_d(n)$ is given in \eqref{hdn},
\begin{align*}
    g_d(n) := \sum_{1\le j\le d-2}\binom{d}{j}
    (-1)^{j-1} (d-1-j)^{j-1}
    \Gamma\left(\tfrac1{d-1-j}\right)^{d-j}
    n^{\frac1{d-1}-\frac{1}{d-1-j}} ,
\end{align*}
and $\eta_d(n) := \phi_{d,d-1}(n) + \phi_{d,d}(n)$.
\end{lmm}
Note that, by \eqref{phi-ddn} and \eqref{phi-dd1n},
\begin{align*}
    \eta_d(n)&= O\left(d^{-1}n^{-d+1-\frac1{d-1}}e^{-n}
    +2^{-d}n^{-d+2+\frac1{d-1}}e^{-2n} \right)\\
    &= O\left(n^{-d+2} e^{-n}\right).
\end{align*}
Also, by the change of variables $t\mapsto v_1\cdots v_j$, we have
\begin{align*}
    f_d(n) &= g_d(n)+h_d(n)+\eta_d(n) \\
    &\qquad+ \sum_{1\le j\le d-2}\binom{d}{j}
    \frac{(-1)^j n^{\frac1{d-1}-\frac{1}{d-1-j}}}{(j-1)!}
    \int_1^\infty t^{-1-\frac1{d-1-j}}(\log t)^{j-1}
    f_{d-j}(nt)\text{d} t,
\end{align*}
which is easier to use for symbolic computation softwares.

We then obtain, for example,
\begin{align*}
    f_3(n) &= 3n^{-\frac12}+ O\left(n^{-\frac32}\right),\\
    f_4(n) &= 4\pi^{\frac32}n^{-\frac16}+O\left(n^{-\frac23}\right),\\
    f_5(n) &= \frac{80\pi^4}{9\Gamma\left(\frac23\right)^4}
    \,n^{-\frac1{12}}-60\pi^{\frac32}n^{-\frac14}
    +O\left(n^{-\frac5{12}}\right).
\end{align*}
But the expressions soon become too messy.

\paragraph{Asymptotic estimate for $g_d(n)$} We derive first a
uniform asymptotic approximation to $g_d(n)$, which will be needed
later. We focus on the case when $d$ tends to infinity with $n$.
\begin{lmm} If $d$ satisfies \eqref{d-rg2}, then
\begin{align}
    g_d(n) &= \frac1{d-1}
    \Gamma\left(\frac1{d-1}\right)^d
    \left\{1-e^{-\rho}+\rho e^{-\rho}\left(\frac{2\rho -1}{2d} +
    \frac{\rho-3}{d^3}\,\log n \right) \right.\nonumber \\
    &\hspace*{4cm}\left. +
    O\left(\frac{\rho e^{-\rho}(\rho^3+1)}{d^2} \left(
    1+\frac{\log^2 n}{d^4}\right)\right)\right\},
    \label{gdn-asymp}
\end{align}
uniformly in $d$.
\end{lmm}
\pf  First, we have
\begin{align*}
    &\frac{\binom{d}{j}(-1)^{j-1} (d-1-j)^{j-1}
    \Gamma\left(\tfrac1{d-1-j}\right)^{d-j}
    n^{-\frac{j}{(d-1)(d-1-j)}}}
    {\frac1{d-1}\Gamma\left(\frac1{d-1}\right)^d}\\
    &\quad= \frac{d^j}{j!}(-1)^{j-1} n^{-\frac j{d^2}}
    \exp\left(-j-\frac{2j^2-j}{2d}- \frac{j(j+2)}{d^3}\log n
    +O\left(\frac{j^3}{d^2}+
    \frac{j^3}{d^4}\log n\right)\right),
\end{align*}
uniformly for $j=o(d^{\frac23})$. Summing over all $j$ gives
\eqref{gdn-asymp}. Here the errors omitted are estimated by the
inequalities
\begin{align*}
    \left\{\begin{array}{rl}
        \binom{d}{j} \!\!\!\!&= O\left(\frac{d^j}{j!}\,
        e^{-\frac{j^2}{2d}}\right),\\
        \Gamma\left(\tfrac1{x}\right)\!\!\!\! &\le x, \qquad(x\ge1)\\
       (d-1-j)^{d-1}\!\!\!\!&\le d^{d-1} e^{-j-\frac{j^2}{2d}},
    \end{array}\right.
\end{align*}
for $1\le j\le d-2$, and we see that the contribution of terms in
$g_d(n)$ with indices larger than, say $j_0 := \lfloor 
d^{\frac35}\rfloor$ are bounded 
above by
\begin{align*}
    &\sum_{j\ge j_0} \binom{d}{j}(-1)^{j-1} (d-1-j)^{j-1}
    \Gamma\left(\tfrac1{d-1-j}\right)^{d-j}
    \, n^{-\frac{j}{(d-1)(d-1-j)}} \\
    &\qquad = O\left(\frac1{d-1}\Gamma\left(\frac1{d-1}\right)^d
    \sum_{j\ge j_0}\frac{\rho^j}{j!}\right)\\
    &\qquad = O\left(\frac1{d-1}\Gamma\left(\frac1{d-1}\right)^d
    \frac{\rho^{j_0}}{j_0!}\right).
\end{align*}
Thus for $d$ in the range \eqref{d-rg2}
\begin{align*}
    j_0\log\rho - \log j_0! &= \tfrac25 d^{\frac35}\log d
    -d^{-\frac75} \log n + d^{\frac35}+O(\log d) \\
    &\le -\left(2^{-\frac 75}-\tfrac1{5}\, 2^{\frac35}\right)
    (\log n)^{\frac3{10}}(\log\log n)^{\frac7{10}}(1+o(1))\\
    &\le -\tfrac3{40} (\log n)^{\frac3{10}}
    (\log\log n)^{\frac7{10}} (1+o(1)),
\end{align*}
so that
\[
    \frac{\rho^{j_0}}{j_0!} = O\left(e^{-\tfrac3{40}
    (\log n)^{\frac3{10}}
    (\log\log n)^{\frac7{10}}(1+o(1))}\right),
\]
and the sum of these terms is asymptotically negligible. The errors
$\sum_{j\ge j_0}\frac{\rho^j}{j!}$ are estimated similarly. 
\qed

\paragraph{Iteration of the $\Phi$-operator} To derive a similar
estimate for $f_d(n)$, we define the operator
\[
    \Phi[f_d](n) := \sum_{1\le j\le d-2}\binom{d}{j}
    (-1)^j n^{\frac1{d-1}-\frac{1}{d-1-j}}
    \int_{(1,\infty)^j}\left(v_1\cdots v_j
    \right)^{-1-\frac1{d-1-j}} f_{d-j}(nv_1\cdots v_j)\text{d}\v{v}.
\]
By iterating the recurrence \eqref{fdn-rr}, we obtain
\begin{align*}
    f_d = g_d +h_d+\eta_d+ \sum_{1\le j\le d-2}
    \Phi^j[g_d+h_d+\eta_d],
\end{align*}
where $\Phi^j[f_d]= \Phi[\Phi^{j-1}[f_d]]$ denotes the $j$-th
iterate of the $\Phi$-operator.

Surprisingly, despite of the complicated forms of the partial sums,
each $\Phi^m[g_d]$ can be explicitly evaluated and differs from
$g_d$ only by a single term.
\begin{lmm} For any $m\ge0$
\begin{align}\label{Phi-m}
    \Phi^m[g_d](n) = \sum_{m< \ell\le d-2}\binom{d}{\ell}
    (-1)^{\ell-1}(d-1-\ell)^{\ell-1}
    \Gamma\left(\tfrac1{d-1-\ell}\right)^{d-\ell}
    n^{\frac1{d-1}-\frac{1}{d-1-\ell}}  \sigma_m(\ell) ,
\end{align}
where $\sigma_m(\ell)$ is always positive and defined by
\begin{align*} 
    \sigma_m(\ell) :=\sum_{\substack{j_1+\cdots+j_{m+1}=\ell\\
    j_1,\dots,j_{m+1}\ge1}}\binom{\ell}{j_1,\cdots,j_{m+1}}.
\end{align*}
\end{lmm}
Note that
\begin{align*}
    \sigma_m(\ell) &= \ell![z^\ell]\left(e^z-1\right)^{m+1}\\
    &=\sum_{1\le r\le m+1}
    \binom{m+1}{r}(-1)^{m+1-r}r^\ell.
\end{align*}

\pf  By definition and by rearranging the terms
\[
    g_d(n) = \sum_{1\le j\le d-2}\binom{d}{j+1}
    (-1)^{d-j} j^{d-2-j} \Gamma\left(\tfrac1{j}\right)^{j+1}
    n^{\frac1{d-1}-\frac{1}{j}}.
\]
Substituting this expression into the $\Phi$-operator, we see that
\begin{align*}
    \Phi[g_d](n) &= \sum_{1\le j\le d-2}\binom{d}{j}
    (-1)^j n^{\frac1{d-1}-\frac{1}{d-1-j}}
    \int_{(1,\infty)^j}\left(v_1\cdots v_j
    \right)^{-1-\frac1{d-1-j}} g_{d-j}(nv_1\cdots v_j)\text{d}\v{v}\\
    &= \sum_{1\le j\le d-2}\binom{d}{j}
    (-1)^j n^{\frac1{d-1}}\\
    &\qquad \times\sum_{1\le \ell\le d-j-2}\binom{d-j}{\ell+1}
    (-1)^{d-j-\ell} \ell^{d-2-j-\ell}
    \Gamma\left(\tfrac1{\ell}\right)^{\ell+1}
    n^{-\frac{1}{\ell}} \int_{(1,\infty)^j}\left(v_1\cdots v_j
    \right)^{-1-\frac1{\ell}}\text{d} \v{v} .
\end{align*}
Then
\begin{align*}
    \Phi[g_d](n)
    &= \sum_{1\le j\le d-2}\binom{d}{j}
    (-1)^j n^{\frac1{d-1}}
    \sum_{1\le \ell\le d-j-2}\binom{d-j}{\ell+1}
    (-1)^{d-j-\ell} \ell^{d-2-\ell}
    \Gamma\left(\tfrac1{\ell}\right)^{\ell+1}
    n^{-\frac{1}{\ell}}  \\
    &= \sum_{1\le \ell\le d-2}\binom{d}{\ell+1}
    (-1)^{d-\ell}\ell^{d-2-\ell}
    \Gamma\left(\tfrac1{\ell}\right)^{\ell+1}
    n^{\frac1{d-1}-\frac{1}{\ell}}
    \sum_{1\le j\le d-2-\ell}\binom{d-1-\ell}{j}\\
    &= \sum_{1\le \ell\le d-2}\binom{d}{\ell+1}
    (-1)^{d-\ell}\ell^{d-2-\ell}
    \Gamma\left(\tfrac1{\ell}\right)^{\ell+1}
    n^{\frac1{d-1}-\frac{1}{\ell}}
    \left(2^{d-1-\ell}-2\right) \\
    &= \sum_{1\le \ell\le d-2}\binom{d}{\ell}
    (-1)^{\ell-1}(d-1-\ell)^{\ell-1}
    \Gamma\left(\tfrac1{d-1-\ell}\right)^{d-\ell}
    n^{\frac1{d-1}-\frac{1}{d-1-\ell}}
    \left(2^{\ell}-2\right) .
\end{align*}
By repeating the same analysis and induction, we prove
\eqref{Phi-m}. \qed

\begin{cor} If $d$ satisfies \eqref{d-rg2}, then
\[
    \Phi^m[g_d](n) \sim (-1)^{m}
    \tfrac1{d-1}\Gamma\left(\tfrac1{d-1}\right)^d
    \left(1-e^{-\rho}\right)^{m+1}\qquad(m=0,1,\dots).
\]
\end{cor}
Summing over all $0\le m\le d-2$, we deduce \eqref{fdn-asymp}
and it remains only the error estimates.

\paragraph{Error analysis} The consideration of $\Phi^m[h_d]$ is
similar and we obtain
\begin{align*}
    \Phi^m[h_d](n) &\le \sum_{m< \ell\le d-2}\binom{d}{\ell}
    2^{-\ell} (d-1-\ell)^{\ell-2}
    \Gamma\left(\tfrac2{d-1-\ell}\right)^{d-\ell}
    n^{\frac1{d-1}-\frac{2}{d-1-\ell}} \sigma_m'(\ell)
\end{align*}
where
\begin{align*}
    \sigma_m'(\ell) &:=\sum_{\substack{j_1+\cdots+j_{m+1}=\ell\\
    j_1,\dots,j_{m+1}\ge1}}\binom{\ell}{j_1,\cdots,j_{m+1}}j_{m+1}\\
    &= \ell![z^\ell]ze^z\left(e^z-1\right)^m\\
    &= \ell \sum_{0\le r\le m}
    \binom{m}{r}(-1)^{m-r}(r+1)^\ell \qquad(m\ge0).
\end{align*}
Thus, with
\[
    \rho _0 := \frac{d}{e n^{2/d^2}}
\]
which is always $\le \log 2$ when $d$ satisfies
\eqref{d-rg2}, we then have
\begin{align*}
    \frac{\Phi^m[h_d](n)}
    {\tfrac{1}{d(d-1)2^d}\Gamma\left(
    \tfrac1{d-1}\right)^d n^{-\frac1{d-1}}}
    &= O\left(\sum_{0\le r\le m}\binom{m}{r}(-1)^{m-r}
    \sum_{\ell\ge0} \frac{\rho_0^\ell}{(\ell-1)!}\,(r+1)^\ell \right)\\
    &= O\left( \rho_0e^{\rho_0}
    \sum_{0\le r\le m}\binom{m}{r}(-1)^{m-r}
    (r+1)e^{r\rho_0}\right)\\
    &= O\left( \rho_0 e^{\rho_0}\left((e^{\rho_0}-1)^{m-1}
    \left((m+1)e^{\rho_0}-1\right) \right)\right).
\end{align*}
Now
\[
    \sum_{0\le m\le d-2} \left((x-1)^{m-1}
    \left((m+1)x-1\right)\right) = O(d^2)
\]
whenever $0\le x\le 2$. It follows that
\[
    \sum_{0\le m\le d-2}\Phi^m[h_d] = O\left(
    2^{-d}d^{-2} \Gamma\left(
    \tfrac1{d-1}\right)^d n^{-\frac1{d-1}}
    \rho_0 e^{\rho_0}\right),
\]
which holds uniformly as long as $e^{\rho_0}\le 2$. This is how the
upper limit of $d$ in \eqref{d-rg2} arises.

In such a case,
\[
    \sum_{0\le m\le d-2}\Phi^m[h_d] = O\left(
    2^{-d}d^{-1} \Gamma\left(
    \tfrac1{d-1}\right)^d n^{-\frac1{d-1}-\frac2{d^2}}
    \right).
\]

We consider now $\Phi^j[\eta_d]$. Note that an exponentially small
term remains exponentially small under the $\Phi$-operator because
\[
    \int_{(1,\infty)^j} (v_1\cdots v_j)^{-1-\alpha}
    e^{-nv_1\cdots v_j} \text{d} {\v{v}} \sim n^{-j} e^{-n}.
\]
So all terms of the forms $\Phi^m[\eta_d]$ are asymptotically
negligible. And we then deduce \eqref{fdn-asymp}. 
\qed

More calculations give
\begin{align*}
    \frac{f_d(n)}{\frac1{d-1}\Gamma\left(\frac1{d-1}\right)^d}
    &= \frac{1-e^{-\rho}}{2-e^{-\rho}} + \frac{\rho
    e^{-\rho}}{(2-e^{-\rho})^3}
    \left(\frac{2\rho -1+(\rho+\tfrac12)e^{-\rho}}{d}\right.\\
    &\quad \left.+
    \frac{2(\rho-3)+\left(\rho+3\right)e^{-\rho}}{d^3}\log n\right)
    +O\left(\frac{\rho e^{-\rho}}{d^2}(\rho^3+1)
    \left(1+\frac{\log^2 n}{d^4}\right)\right).
\end{align*}

Note that the range \eqref{d-rg1} arises because we had to drop
factors of the form $(-1)^j$ in estimating the sum of $h_d(n)$. With
a more careful analysis along the same inductive line, we can extend
the range of uniformity of \eqref{fdn-asymp}. 

\end{document}